\documentclass[12pt,a4paper]{article}
\usepackage[utf8]{inputenc}
\usepackage{amsmath} 
\usepackage{amsfonts}
\usepackage{amssymb}
\usepackage{amsthm}
\usepackage{mathtools}
\usepackage{enumerate}
\usepackage{paralist}
\usepackage{istgame}

\newcommand{\pd}{\partial}
\usepackage{tikz}
\usepackage{float}
\usepackage{makecell}

\usepackage[colorlinks,
            linkcolor=blue,
            anchorcolor=blue,
            citecolor=blue,
            urlcolor=blue
            ]{hyperref}
\usepackage[capitalize,nameinlink,noabbrev]{cleveref}
\crefname{ineq}{Inequality}{inequalities}
\creflabelformat{ineq}{#2{\upshape#1}#3} 

\usepackage{autonum}

\usepackage{apptools}
\crefname{subappendix}{\IfAppendix{section}{appendix}}{\IfAppendix{sections}{appendices}s}
\creflabelformat{equation}{#2\textup{#1}#3}

\usepackage{times}

\usepackage{lipsum}
\usepackage{newpxtext,newpxmath}

\usepackage{natbib}
\usepackage{etoolbox}

\makeatletter
\newcommand\bibstyle@comma{\bibpunct(),a,,}
\newcommand\bibstyle@semicolon{\bibpunct();a,,}
\makeatother

\pretocmd\citet{\citestyle{comma}}\relax\relax
\pretocmd\Citet{\citestyle{comma}}\relax\relax
\pretocmd\citep{\citestyle{semicolon}}\relax\relax
\pretocmd\Citep{\citestyle{semicolon}}\relax\relax

\usepackage{geometry}
\usepackage{setspace} 

\usepackage{forloop}
\newcommand{\defcal}[1]{\expandafter\newcommand\csname
c#1\endcsname{{\mathcal{#1}}}}
\newcommand{\defbb}[1]{\expandafter\newcommand\csname
bb#1\endcsname{{\mathbb{#1}}}}
\newcommand{\defbf}[1]{\expandafter\newcommand\csname
bf#1\endcsname{{\mathbf{#1}}}}
\newcounter{calBbCounter}
\forLoop{1}{26}{calBbCounter}{
	\edef\letter{\Alph{calBbCounter}}
	\expandafter\defcal\letter
	\expandafter\defbb\letter
	\expandafter\defbf\letter
}

\newcommand{\rnum}[1]{\lowercase\expandafter{\romannumeral #1\relax}}
\newcommand{\Rnum}[1]{\uppercase\expandafter{\romannumeral #1\relax}}

\usepackage{accents}

\newtheorem{theorem}{Theorem}
\newtheorem{prop}[theorem]{Proposition}
\newtheorem{lemma}{Theorem}

\newtheorem{Corollary}{Theorem}
\newtheorem{coro}[Corollary]{Corollary}
\newtheorem{lem}[lemma]{Lemma}
\theoremstyle{definition}

\newtheorem{assumption}{Theorem}
\newtheorem{as}[assumption]{Assumption}

\usepackage{environ}
\NewEnviron{iproof}[1][Proof]{\begin{proof}[\bfseries #1]\BODY\end{proof}}{}

\usepackage{epigraph}

\usepackage{appendix}

\usepackage{apptools}
\AtAppendix{\counterwithin{lemma}{section}}
\AtAppendix{\counterwithin{theorem}{section}}

\title{Transparency and Policymaking with Endogenous Information Provision}
\author{Hanzhe Li\footnote{The University of Hong Kong; hanzhe.li@connect.hku.hk. The article is based on my MPhil thesis (2021) at the Chinese University of Hong Kong. I am very grateful for the advice and encouragement of Jimmy Chan and Jin Li, and thank Yifei Dai, Antoni-Italo De Moragas, Yue Han, Wei He, Duozhe Li, Wing Suen, Yanhui Wu, and Xinbei Zhou for their valuable suggestions. The article is also benefited greatly from the suggestions and comments of the editor and three anonymous referees.}}

\date{\today}

\begin{document}

\doublespacing
\maketitle

\begin{abstract}
    \noindent How does the politician's reputation concern affect information provision when the information is endogenously provided by a biased lobbyist? I develop a model to study this problem and show that the answer depends on the transparency design. When the lobbyist’s preference is publicly known, the politician's reputation concern induces the lobbyist to provide more information. When the lobbyist’s preference is unknown, the politician's reputation concern may induce the lobbyist to provide less information. One implication of the result is that given transparent preferences, the transparency of decision consequences can impede information provision by moderating the politician's reputational incentive.\bigskip

    \bigskip
\end{abstract}

\newcommand{\chapquote}[3]{\begin{quotation} \textit{#1} \end{quotation} \begin{flushright} - #2, \textit{#3}\end{flushright} }

\chapquote{``Whenever you do a thing, act as if all the world were watching.''}{Thomas Jefferson}{1785}

\section{Introduction}\label{I}
Politicians have the desire to appear competent. One prominent example is congress members, who actively seek out opportunities to enhance their reputation, with their congressional offices serving as vehicles for reputation-building ~\citep{fenno1973congressmen, salisbury1981us}. This raises the concern that politicians may prioritize accumulating reputation over making the correct decision, foregrounding the reputation-concern problem studied by economists (e.g., \citealp{prendergast1996impetuous}, \citealp{10.2307/2669237}, and \citealp{10.2307/4135242}).

A common way to tackle the reputation-concern problem is to change transparency designs on politicians. This direction has been widely explored in the literature, which focuses on the strategic motives of politicians and mostly assumes exogenous information for policymaking (see \citealp{Fox2007}, \citealp{levy2007decision}, \citealp{stasavage2007}, and \citealp{ashworth2010does} among others).\footnote{As exceptions, \citet{bar2012transparency} studies how transparency affects effort in acquiring information; \citet{salas2019persuading} studies the public persuasion of a career-concerned politician by assuming that the politician does not know her ability, which is more related to \citet{kolotilin2018optimal} and does not address transparency issues.} This literature, however, neglects the fact that policy-relevant information is often endogenously provided by biased lobbyists \citep{baumgartner2009lobbying}. Hence, this article endogenizes information provision as a Bayesian persuasion game \citep{kamenica2011bayesian} to shed some new light. In particular, I answer the following two questions: How does the politician's reputation concern affect information provision when a biased lobbyist endogenously provides the information? What is the role of transparency during lobbying and policymaking?

In the model, the lobbyist (he) privately commits to an experiment to influence the politician (she). If the politician is high-ability, she cannot be influenced as she knows the correct action that matches the underlying state. But if the politician is low-ability, she relies on the lobbyist for information. The politician seeks to take the correct action and be perceived as high-ability. The lobbyist, being biased, prefers one certain action regardless of the state. The public's belief about the politician's ability is called the politician's reputation. Before updating their belief, the public observes the politician's action and the decision consequence (i.e., whether the action is correct).

The main result is as follows. When the lobbyist's preference is publicly known, the politician's reputation concern incentivizes her to take the lobbyist’s disliked action. This incentive induces the lobbyist to provide more information. As a result, reputation concerns help politicians make better decisions when the lobbyist's preference is transparent. In contrast, when the lobbyist's preference is unknown to the public, the politician's reputation concern may induce the lobbyist to provide less information. 

To illustrate the intuition, suppose that the lobbyist recommends his preferred action. A low-ability politician follows the recommendation because she is not otherwise informed about the state, but a high-ability politician observes the state and may go against the recommendation. If the public sees the politician take the lobbyist's disliked action, it is inferred that she is more likely to be high-ability. This inference provides a reputational incentive to take the lobbyist's disliked action when the lobbyist's preference is publicly known. Consequently, to persuade the reputation-concerned politician, the lobbyist needs to provide more information.

When the lobbyist’s preference is unknown, the public does not know the lobbyist's disliked action. They can only perceive one action as lobbyist-disliked, which may coincide or differ from the actual disliked action. When the perceived disliked and actual disliked actions coincide, the argument in the previous paragraph implies that the politician is incentivized to take the actual disliked action. When the perceived disliked and actual disliked actions differ, however, the politician's reputation concern only incentivizes her to take the perceived disliked action. This causes the lobbyist to provide less information to persuade the politician.

The intuition suggests that when the lobbyist's preference is known, the transparency of decision consequences can impede information provision by moderating the politician's reputational incentive. Specifically, the moderating effect arises because when the decision consequence is transparent, the politician has to be correct to gain a positive reputation. This requirement moderates the reputational incentive to take the lobbyist's disliked action. Therefore, from nontransparent to transparent consequences, the lobbyist can persuade the politician with less information provision.

These findings have practical implications for the design of lobby registers. Many countries, including the US, only require mandatory disclosure of lobbyists' identity and their lobbying targets \citep{/content/publication/c6d8eff8-en}. But this information is not sufficient to infer lobbyists' preferences. For example, existing firms may lobby against regulations to reduce their costs, but they may also lobby for the same regulations to prevent potential competitors. In 2019, the European Parliament adopted a lobby register that requires lobbyists to disclose their policy preferences. This article provides a rationale for such disclosure. The disclosure of policy preferences can improve the information provided by lobbyists and help politicians formulate appropriate policies.

The findings also highlight a novel conflict between the conventional transparency of decision consequences and the new transparency of lobbyists' preferences. When a politician faces a decision problem but has to rely on external information sources, a subsequent public evaluation of the politician's performance can be socially beneficial only if the intent behind the information input is difficult to reveal. If the intent is readily transparent, the public may be better off not knowing the politician's performance as the external information source will have to provide more information \emph{ex ante}.

The remaining part of this section discusses the related literature. \cref{M} presents the baseline model. \cref{MR} characterizes the equilibrium experiment and shows the effect of the politician's reputation concern. \cref{LG} studies the nontransparency of the lobbyist's preference. \cref{TC} studies the nontransparency of decision consequences. \cref{D} considers other transparency forms. \cref{C} concludes. Proofs can be found in \cref{App1}.\bigskip

\noindent\textbf{Literature Review.} Starting from \citet{holmstrom1982} and \citet{holmstrom1999managerial}, a large literature studies how reputation concerns (or career concerns) provide incentives. In the setting with moral hazard, \citet{SUURMOND2004} show that the agent's reputation concern incentivizes him to acquire more information. \citet{stepanov2020biased} shows that when an agent has career concerns, an evaluator against the agent can make him put in more effort. Similarly, in a hierarchy, the leader's reputation concern makes her conservative, and thus the bottom specialist has to make more effort to get his project approved \citep{Fu2022}. My article differs from the previous studies in the information component that transmits from the lobbyist to the politician.

The information component connects this article to the reputational cheap-talk literature (e.g., \citealp{morris2001political}, \citealp{gentzkow2006media}, \citealp{ottaviani2006professional, ottaviani2006reputational}, \citealp{alizamir2020warning}, and \citealp{balmaceda2021private}). That literature studies sender-receiver games and focuses on senders' reputation concerns. However, this article focuses on the reputation concern of decision-makers.

This article studies transparency designs and is closely related to articles about nontransparent preferences. Endogenous information provision and the focus on the lobbyist's bias distinguish this article from \citet{DEMORAGAS2022104282}, which shows the benefit of hiding the politician's bias. In cheap-talk games, \citet{li2008mandatory} show that nondisclosure of the lobbyist's bias can benefit the public by making the lobbyist more credible and facilitating more precise communication. The channel, however, is blocked by the commitment assumption in this article, where the transparency of the lobbyist's preference matters only for the reputation channel.

This article is also related to the literature on transparency and reputation concerns. \citet{prat2005wrong} and \citet{FU201415} show that it is socially beneficial to reveal decision consequences because it disciplines the politician by moderating her reputational incentive. \citet{fox2012costly} derive similar results but also show the opposite possibility when the cost between incorrect actions is asymmetric. This article differs from them in the endogenously provided information. Although the moderating effect of revealing decision consequences is ex-post beneficial to the public, it is ex-ante harmful because it diminishes the lobbyist's incentive to provide more information. Such inconsistency is reminiscent of the trade-off that arises between ex-ante incentives and ex-post efficiencies in other endogenous information problems (e.g., \citealp{che2009opinions} and \citealp{frankel2019improving}). \citet{acharya2021building} study how the government builds a reputation for trustworthiness by handling privately observed crises. \citet{Lifei2023} show that in a society, seeing positive outcomes of crisis management in other societies may boost optimism and block appropriate policies.

\section{Baseline Model}\label{M}
The process of lobbying and policymaking is modeled as a persuasion game between the lobbyist (he) and the politician (she). The lobbyist is biased toward $a$ with probability $\alpha$ (``lobbyist A'') and biased toward $b$ with probability $1-\alpha$ (``lobbyist B''). The politician is high-ability with probability $\tau\in(0,1)$ and low-ability with probability $1-\tau$. Both the bias and the ability are types drawn before the persuasion game. Although the ability is private information, the bias is public information.

The persuasion game begins with the lobbyist committing to an experiment,\footnote{The commitment may reflect the repeated interaction between the lobbyist and the politician (see \citealp{groll2017repeated} for relevant background). When they interact repeatedly, the equilibrium outcome can replicate the optimal persuasion with commitment power ~\citep{kolotilin2021,che2022keeping}.} $\sigma:\Omega\to\Delta(S)$, where $\Omega=\{A,B\}$ denotes the state space and $S=\{\Tilde{a},\Tilde{b}\}$ denotes the recommendation space.\footnote{A binary recommendation space is sufficient because the action is binary, and the low-ability politician takes a pure action after receiving any recommendation in equilibrium (\cref{lem no randomization}).} Then, the state $\omega\in\Omega$ realizes according to the prior $\mu_0:=\bfP(\omega=A)\in(0,1/2)$, and the politician receives a recommendation $s\in S$ with probability $\sigma(s|\omega)$. Finally, observing the experiment and the recommendation, the politician takes an action from $X=\{a,b\}$. Her strategy, $\gamma(\sigma,s,z)\in\Delta(X)$, specifies a possibly randomized action for each combination of experiments, recommendations, and private information about the state ($z=\omega$ for the high-ability politician and $z=\emptyset$ for the low-ability politician).

After the persuasion game, the public updates their belief about the politician's ability. The public observes the politician's action, the decision consequence (i.e., the correctness of the action---whether $(x,\omega)\in\{(a,A),(b,B)\}$), and the lobbyist's preference. The public does not observe the recommendation, the experiment, or the politician's strategy but has consistent beliefs about them in equilibrium. This transparency setting about what the public can observe distinguishes the baseline model, which will be studied in \cref{MR}, from other transparency settings that \cref{LG}, \ref{TC}, and \ref{D} consider. The public's belief of the politician being high-ability is called the politician's reputation. Denote it as $\tau(x,\omega)$ for $x\in X$ and $\omega\in\Omega$.

Both the lobbyist and the politician are risk neutral. The lobbyist's utility is type-specific: $u^L(x,\omega)$ equals $\mathbf{1}_{x=a}$ for lobbyist A and $\mathbf{1}_{x=b}$ for lobbyist B. The politician's utility is given by $$u^P(x,\omega)=\mathbf{1}_{(x,\omega)\in\{(a,A),(b,B)\}}+\theta\tau(x,\omega),$$ where the indicator function captures the decision quality, and $\theta>0$ shows the intensity of the politician's reputation concern. For simplicity, the reputational payoff is assumed to be linear, but the form can be relaxed as any strictly increasing function. The public's welfare is defined as $\bfE[\mathbf{1}_{(x,\omega)\in\{(a,A),(b,B)\}}]$.

Throughout the article, $\pi$ denotes lobbyist A's experiment, and $\rho$ denotes lobbyist B's experiment. An experiment $\sigma$ is said to be more informative than $\sigma'$ if $\sigma$ Blackwell dominates $\sigma'$ ~\citep{blackwell1951equivalent,blackwell1953equivalent}, and the order is denoted as $\sigma\geq\sigma'$.

The solution concept is Perfect Bayesian Equilibrium. An equilibrium is a profile $(\pi,\rho,\gamma)$ that satisfies the following: (1) the public has correct conjectures about these strategies and generates the politician's reputation accordingly, (2) $\gamma$ specifies the politician's optimal action upon any information set, and (3) $\pi$ and $\rho$ are the best responses of different types of lobbyists. I eliminate all equilibria in which the high-ability politician may make incorrect decisions. Hence, in equilibrium, $(\gamma(\sigma, s,\omega),\omega)\in\{(a,A),(b,B)\}$ for any experiment $\sigma$, $s\in S$, and $\omega\in\Omega$. In line with the standard refinements (e.g., \citealp{levy2004anti}, \citealp{prat2005wrong}, \citealp{fox2012costly}, and \citealp{FU201415}), my refinement rules out perverse equilibria in which the high-ability politician may signal her ability by deliberately taking incorrect actions.

\section{Basic Results}\label{MR}
This section characterizes the equilibrium in the baseline model, where the public observes the politician's action, the decision consequence, and the lobbyist's preference. I first describe the equilibrium experiment when the politician has no reputation concern ($\theta=0$) and then move on to the baseline model where the politician strictly values her reputation ($\theta>0$). Finally, I show the effect of the politician's reputation concern.

Let $\mu_s:=\mathbf{P}(\omega=A|s)$ denote the politician's posterior about the state conditional on some recommendation $s$. The posterior can be calculated following the experiment design and Bayes' rule:
$$\mu_s=\frac{\mu_0\sigma(s|A)}{\mu_0\sigma(s|A)+(1-\mu_0)\sigma(s|B)} \text{ for }s\in\{\Tilde{a},\Tilde{b}\}.$$
Note that $\bfE_s[\mu_s]=\mu_0$. Conversely, the posteriors can pin down the experiment design through these formulas. We may take either way to reason about experiments.

Take lobbyist A as an example. He targets the low-ability politician because the high-ability politician observes the state and takes the correct action. When the politician has no reputation concern ($\theta=0$), the lobbyist's problem is a canonical example in the Bayesian persuasion literature. He aims to design an experiment that maximizes the probability of recommending action $a$ subject to an obedience constraint that it must be optimal for the low-ability politician to follow the lobbyist's recommendation. \citet{kamenica2011bayesian} show that the solution, denoted by $\pi^0$, sets $\pi^0(\tilde{a}|A)=1$ and $\pi^0(\Tilde{a}|B)$ such that the low-ability politician is indifferent between $a$ and $b$ after receiving $\Tilde{a}$. This optimal experiment splits the prior $\mu_0$ into $\mu^0_{\Tilde{a}}=1/2$ and $\mu^0_{\Tilde{b}}=0$, upon which the low-ability politician takes $a$ or $b$, respectively.

Then, the politician gains her reputation through the action and the decision consequence. If the politician takes an incorrect action, she is revealed as low-ability. Formally, $\tau(a,B)=\tau(b,A)=0$.\footnote{If an incorrect action happens off the equilibrium path (e.g., given $\pi^0$, taking $b$ in $A$ is off-path), the zero reputation can be justified as the limit of equilibrium reputations in the games where the lobbyist only has an imperfect signal about the state, but it converges to being perfect.}
If the politician takes the correct action, she gains a positive reputation:
\begin{equation}
    \tau^0(a,A)=\tau,\ \tau^0(b,B)=\frac{\tau}{\tau+(1-\tau)\pi^0(\Tilde{b}|B)}.
\end{equation}
Note $\tau^0(a,A)<\tau^0(b,B).$ This inequality can be reasoned conditional on $\Tilde{a}$ because both high-ability and low-ability politicians take $b$ upon $\mu^0_{\Tilde{b}}=0$. When the lobbyist sends $\Tilde{a}$, a low-ability politician takes $a$, but a high-ability politician may take $b$. Thus, the politician is more likely to be high-ability when she takes $b$ than when she takes $a$.

Now, suppose that the politician strictly values her reputation ($\theta>0$), and lobbyist A still uses $\pi^0$. In addition, assume that the low-ability politician is obedient, as this will be shown to hold in equilibrium. Given $\Tilde{a}$, the politician's reputational payoff from $a$ is $\theta\mu^0_{\Tilde{a}}\tau^0(a,A)$, and that from $b$ is $\theta(1-\mu^0_{\Tilde{a}})\tau^0(b,B)$. The reputation loss from taking $a$ instead of $b$ is $\theta[(1-\mu^0_{\Tilde{a}})\tau^0(b,B)-\mu^0_{\Tilde{a}}\tau^0(a,A)]$. The politician's quality gain from taking $a$ instead of $b$ is $2\mu^0_{\Tilde{a}}-1$. Because $\mu^0_{\Tilde{a}}=1/2$, the reputation loss is larger than the quality gain:
\begin{equation}
    2\mu^0_{\Tilde{a}}-1=0<\theta[(1-\mu^0_{\Tilde{a}})\tau^0(b,B)-\mu^0_{\Tilde{a}}\tau^0(a,A)].
\end{equation}
Hence, it is optimal for the low-ability politician to take $b$ after receiving $\Tilde{a}$. As $\pi^0$ does not satisfy the obedience constraint, we can conclude from the following lemma and corollary that $\pi^0$ is not lobbyist A's equilibrium experiment when $\theta>0$.

\begin{lem}\label{lem no randomization}
There exists no equilibrium experiment with which the low-ability politician takes a randomized action after receiving some recommendation.\footnote{The two recommendations $\Tilde{a}$ and $\Tilde{b}$ are understood as identical recommendations when $\mu_{\Tilde{a}}=\mu_{\Tilde{b}}$.}
\end{lem}

The proof of the lemma shows that if the low-ability politician randomizes her action after receiving some recommendation, the lobbyist can improve his payoff with a new experiment. Hence, the politician only takes pure (behavioral) actions on the path of any equilibrium. According to the proof, the result holds both for the baseline model and the transparency settings in later sections. 

The lemma has an immediate corollary, which specifies the behavior of the low-ability politician given the equilibrium experiment that she faces. An equilibrium experiment is called \emph{essentially uninformative} if it can be replaced with an uninformative experiment without affecting the equilibrium outcome (the joint distribution of the state and the politician's action).

\begin{coro}\label{coro obedience}
    For any lobbyist, (\rnum{1}) if the equilibrium experiment is essentially uninformative, the low-ability politician takes the lobbyist's preferred action after receiving any recommendation, or (\rnum{2}) if the equilibrium experiment is not essentially uninformative, the low-ability politician takes $a$ after receiving $\Tilde{a}$ and $b$ after receiving $\Tilde{b}$.
\end{coro}

According to the corollary, the low-ability politician's behavior has two possible patterns. When the equilibrium experiment she faces can be uninformative, she always takes the lobbyist's preferred action (the politician always knows the lobbyist's preference from the equilibrium experiment). When the equilibrium experiment must be informative, she takes whatever is recommended.

Consider the first possibility for lobbyist A. If the low-ability politician always takes $a$, the public infers that the politician is high-ability if she takes $b$. Then, as $\mu_{\Tilde{b}}\leq\mu_0<1/2$, it is optimal for the low-ability politician to take $b$ after receiving $\Tilde{b}$. This contradiction shows that for lobbyist A, the low-ability politician takes $a$ only after receiving $\Tilde{a}$. Therefore, similar to the no reputation concern case, lobbyist A aims to maximize the probability of sending $\Tilde{a}$ subject to the obedience constraint.

To achieve the goal, lobbyist A must send $\Tilde{a}$ in $A$ so that $\pi(\Tilde{a}|A)=1$ and make the obedience constraint binding after sending $\Tilde{a}$. This gives rise to the following three equations, which characterize lobbyist A's equilibrium experiment, $\pi^*$ (note that $\pi(\Tilde{a}|\omega)+\pi(\Tilde{b}|\omega)=1$ for any $\omega$):
\begin{equation}\label{eq basic posterior}
   \text{(Posteriors)}\quad\mu^*_{\Tilde{a}}=\frac{\mu_0}{\mu_0+(1-\mu_0)\pi^*(\Tilde{a}|B)},\  \mu^*_{\Tilde{b}}=0.
\end{equation}
\begin{equation}\label{eq basic reputation}
    \text{(Reputations)}\quad\tau^*(a,A)=\tau,\ \tau^*(b,B)=\frac{\tau}{\tau+(1-\tau)\pi^*(\Tilde{b}|B)}.
\end{equation}
\begin{equation}\label{eq basic equality}
    \text{(The binding obedience constraint)}\quad  2\mu^*_{\Tilde{a}}-1=\theta[(1-\mu^*_{\Tilde{a}})\tau^*(b,B)-\mu^*_{\Tilde{a}}\tau^*(a,A)].
\end{equation}
The last equation implies that the low-ability politician is indifferent between $a$ and $b$ upon $\mu^*_{\Tilde{a}}$. Further, it is optimal for her to take $a$ after receiving $\Tilde{a}$ and take $b$ after receiving $\Tilde{b}$. This strategy is reflected in the politician's reputation. 

Similarly, lobbyist B's equilibrium experiment, $\rho^*$, is characterized by:
\begin{equation}\label{eq LB basic posterior}
   \text{(Posteriors)}\quad\mu^*_{\Tilde{a}}=1,\  \mu^*_{\Tilde{b}}=\frac{\mu_0\rho^*(\Tilde{b}|A)}{\mu_0\rho^*(\Tilde{b}|A)+1-\mu_0}.
\end{equation}
\begin{equation}\label{eq LB basic reputation}
    \text{(Reputations)}\quad\tau^*(a,A)=\frac{\tau}{\tau+(1-\tau)\rho^*(\Tilde{a}|A)},\ \tau^*(b,B)=\tau.
\end{equation}
\begin{equation}\label{eq LB basic equality}
    \text{(The binding obedience constraint)}\quad  2\mu^*_{\Tilde{b}}-1=\theta[(1-\mu^*_{\Tilde{b}})\tau^*(b,B)-\mu^*_{\Tilde{b}}\tau^*(a,A)].
\end{equation}
However, because $\mu_0<1/2$, these equations may not admit a solution with $\mu^*_{\Tilde{b}}\in(0,\mu_0)$. If the prior reputation $\tau$ is high enough, or if the politician does not care enough about her reputation (i.e., $\theta$ is small), it may be optimal for the low-ability politician to take $b$ even though $\rho^*$ is uninformative. In that case, the obedience constraint need not be binding. Let $\underline{\theta}$ denote the value of $\theta>0$ such that \cref{eq LB basic equality} holds with $\rho^*(\Tilde{a}|A)=0$ and $\mu^*_{\Tilde{b}}=\mu_0$. This positive threshold exists if $\tau<\mu_0/(1-\mu_0)$. 

The next lemma summarizes the result.

\begin{lem}\label{lem equilibrium}
In the baseline model:
\begin{enumerate}[(i)]
    \item Lobbyist A's equilibrium experiment $\pi^*$ is unique and determined by \cref{eq basic posterior}, \ref{eq basic reputation}, and \ref{eq basic equality}. In particular, $\mu^*_{\Tilde{a}}>1/2$.
    \item Lobbyist B's equilibrium experiment $\rho^*$ is informative if $\tau<\mu_0/(1-\mu_0)$ and $\theta>\underline{\theta}$, and essentially uninformative otherwise. When $\rho^*$ must be informative, it is unique and determined by \cref{eq LB basic posterior}, \ref{eq LB basic reputation}, and \ref{eq LB basic equality}.
\end{enumerate}
\end{lem}

The lemma shows that biased lobbyists must provide (weakly) more information when the politician has reputation concerns than when she does not. For lobbyist A,  $\pi^*$ is more informative than $\pi^0$ as $\mu^*_{\tilde{a}}>1/2$. For lobbyist B, even though $\mu_0<1/2$, $\rho^*$ has to be informative if the prior reputation is low and the politician has a strong enough reputation concern. 

This result is foreshadowed by the analysis of the no reputation concern case, which indicates that the lobbyist's disliked action better signals high ability than the lobbyist's preferred action (e.g., $\tau^0(b,B)>\tau^0(a,A)$ for lobbyist A). This reputational difference can incentivize the politician to take the lobbyist's disliked action if the politician has reputation concerns. In turn, the lobbyist provides more information to persuade the politician to take his preferred action.

\begin{prop}\label{prop beneficial}
In the baseline model, an increase in $\theta$ makes $\pi^*$ and $\rho^*$ more informative and increases the public's welfare.
\end{prop}

The proposition shows that the politician's reputation concern improves information provision in the baseline model. Specifically, it incentivizes the politician to take the lobbyist's disliked action, which leads the lobbyist to provide more information for the persuasion of the politician. As the low-ability politician receives more information, she makes a better decision, improving the public's welfare.

\citet{levy2004anti} shows that the politician's reputation concern is harmful to the public because it leads to inefficient use of information. For example, even if the posterior satisfies $\mu>1/2$, the politician takes $b$ as long as $\tau(b,B)$ is larger enough than $\tau(a,A)$, showing a reputational incentive to go against the posterior. However, this reputational incentive produces a different result when a lobbyist controls the information environment. The lobbyist will provide enough information to counteract this reputational incentive, and thus, information will be efficiently used. A stronger reputational incentive only disciplines the lobbyist to provide more information. 

For this result, the lobbyist must be free or at least not too restricted in designing the experiment. To see why, suppose that the posterior $\mu_{\Tilde{a}}$ induced by $\pi^*$ cannot exceed $\Bar{\mu}:=\mu^*_{\Tilde{a}}|_{\theta=1}$. Then, given $\theta>1$, lobbyist A cannot design an experiment to satisfy the obedience constraint. Instead, she uses $\pi^*|_{\theta=1}$, and the equilibrium is determined by:
\begin{equation}
    \mu^*_{\Tilde{a}}=\Bar{\mu}=\frac{\mu_0}{\mu_0+(1-\mu_0)\pi^*(\Tilde{a}|B)},\ \mu^*_{\Tilde{b}}=0,
\end{equation}
\begin{equation}
    \tau^*(a,A)=\frac{\tau}{\tau+(1-\tau)p},\ \tau^*(b,B)=\frac{\tau}{\tau+(1-\tau)[\pi^*(\Tilde{a}|B)(1-p)+\pi^*(\Tilde{b}|B)]},
\end{equation}
and \cref{eq basic equality}. Notice that the low-ability politician takes $b$ with probability $1-p$ after receiving $\Tilde{a}$. By doing so, she sacrifices the higher quality of action $a$ for the higher reputation of action $b$. This behavior is inefficient in information use and is suboptimal for the public's welfare. However, if the low-ability politician has stronger reputation concerns, she is more inclined to take $b$. The probability $1-p$ will increase as $\theta$ increases. As a result, the politician's reputation concern harms the public's welfare. The lack of endogenous information provision brings back the result in \citet{levy2004anti}.

\section{Nontransparency of the Lobbyist's Preference}\label{LG}
This section studies the nontransparency of the lobbyist's preference: how it affects information provision and changes the effect of the politician's reputation concern. In this section, the public does not know the lobbyist's preference but knows its distribution. The public also knows the action taken by the politician and the decision consequence. 

In this setup, the equilibrium experiment is denoted as $\pi^{**}$ for lobbyist A and $\rho^{**}$ for lobbyist B. Recall that the politician observes the experiment she faces. Even though the lobbyist's preference is nontransparent, the politician can know the lobbyist's preference from the equilibrium experiment. 

\begin{lem}\label{lem experiments nontransparent preferences}
    Suppose that the lobbyist's preference is unknown to the public.
    \begin{enumerate}[(i)]
        \item If $2\mu_0-1<\theta[(1-\mu_0)\tau/(\tau+(1-\tau)(1-\alpha))-\mu_0\tau/(\tau+(1-\tau)\alpha)]$, $\pi^{**}$ is unique and informative, but $\rho^{**}$ is essentially uninformative.
        \item If $2\mu_0-1=\theta[(1-\mu_0)\tau/(\tau+(1-\tau)(1-\alpha))-\mu_0\tau/(\tau+(1-\tau)\alpha)]$, both $\pi^{**}$ and $\rho^{**}$ are uninformative.
        \item If $2\mu_0-1>\theta[(1-\mu_0)\tau/(\tau+(1-\tau)(1-\alpha))-\mu_0\tau/(\tau+(1-\tau)\alpha)]$, $\rho^{**}$ is unique and informative, but $\pi^{**}$ is essentially uninformative.
    \end{enumerate}
\end{lem}

The lemma shows that one of $\pi^{**}$ and $\rho^{**}$ is essentially uninformative. When one type of lobbyist designs an informative experiment, it is optimal for him to make the low-ability politician indifferent between $a$ and $b$ after his preferred action is recommended. For lobbyist A, this is written as
\begin{equation}\label{eq TP A equality}
    \text{(The binding obedience constraint)}\quad 2\mu^{**}_{\Tilde{a}}-1=\theta[(1-\mu^{**}_{\Tilde{a}})\tau^{**}(b,B)-\mu^{**}_{\Tilde{a}}\tau^{**}(a,A)],
\end{equation}
and the existence of the solution (i.e., $\pi^{**}$) is determined by the comparison between $2\mu_0-1$ and $\theta[(1-\mu_0)\tau/(\tau+(1-\tau)(1-\alpha))-\mu_0\tau/(\tau+(1-\tau)\alpha)]$. 

Suppose that $\pi^{**}$ exists and is informative. Notice that $\tau^{**}(a,A)$ and $\tau^{**}(b,B)$ are the same across different types of lobbyists. Then, by designing an uninformative experiment, lobbyist B can tilt \cref{eq TP A equality} as $2\mu_0-1<\theta[(1-\mu_0)\tau^{**}(b,B)-\mu_0\tau^{**}(a,A)]$ and persuade the low-ability politician always to take $b$. This implies that lobbyist B will provide no information if lobbyist A's equilibrium experiment is informative. Similarly, lobbyist A will provide no information if lobbyist B's equilibrium experiment is informative. 

It is worth noting that the lobbyist wants to participate in the game even if he does not provide information after participating. For example, consider part (\rnum{1}) of \cref{lem experiments nontransparent preferences}, where lobbyist B does not provide information in equilibrium. If, in addition, he does not participate, and if this is publicly observed,\footnote{If the public does not observe whether the lobbyist participates, it can be optimal not to participate for the lobbyist who does not provide information in equilibrium. In that case, the prior over that lobbyist type can be replaced with the possibility of no lobbyist.} the low-ability politician may take $a$ to signal that she has private information beyond the prior $\mu_0<1/2$. But if lobbyist B participates, the low-ability politician only takes $b$. Therefore, it is optimal for lobbyist B to participate in the game. The equilibrium feature---the lobbyist lobbies the low-ability politician even though the politician will take the lobbyist's preferred action without receiving any information---resembles the ``friendly lobbying'' phenomenon, in which lobbyists target politicians who already support lobbyist-preferred positions (e.g., \citealp{austen-smith1994}, \citealp{schnakenberg2017}, and \citealp{awad2020}).

In the following, if the equilibrium experiment is essentially uninformative, I take it as uninformative to facilitate the comparative statics. Thus, both $\pi^{**}$ and $\rho^{**}$ are unique. The next proposition compares $\pi^{**}$ and $\rho^{**}$ with the equilibrium experiments $\pi^*$ and $\rho^*$ in the transparent case.

\begin{prop}\label{prop LG comparative}
The nontransparency of the lobbyist's preference, compared with the transparency, decreases information provision and the public's welfare: $\pi^{**}<\pi^{*}$ and $\rho^{**}\leq \rho^{*}$. 
\end{prop}

The proposition shows that the nontransparency of the lobbyist's preference impedes information provision. The result arises because, unlike in the transparent case, actions are not known as lobbyist-preferred or lobbyist-disliked. This uncertainty makes the public less determined about which action signals high ability better than the other and weakens the politician's incentive to take the lobbyist's disliked action. As a result, the lobbyist can provide less information to persuade the politician.

Take lobbyist A as an example. If $\pi^{**}$ is uninformative, it is clear that lobbyist A provides less information than $\pi^*$. Then, suppose that $\pi^{**}$ is informative. According to the proof of \cref{lem experiments nontransparent preferences}, $\pi^{**}$ is characterized by
\begin{equation}\label{eq TP A posteriors}
    \text{(Posteriors)}\quad\mu^{**}_{\Tilde{a}}=\frac{\mu_0}{\mu_0+(1-\mu_0)\pi^{**}(\Tilde{a}|B)},\  \mu^{**}_{\Tilde{b}}=0,
\end{equation}
\begin{equation}\label{eq TP A reputations}
    \text{(Reputations)}\quad\tau^{**}(a,A)=\frac{\tau}{\tau+(1-\tau)\alpha},\  \tau^{**}(b,B)=\frac{\tau}{\tau+(1-\tau)[\alpha\pi^{**}(\Tilde{b}|B)+(1-\alpha)]},
\end{equation}
and \cref{eq TP A equality}, whereas $\rho^{**}$ is uninformative. In equilibrium, the low-ability politician takes $a$ after receiving $\Tilde{a}$ from lobbyist A and takes $b$ otherwise (the politician always knows the lobbyist's preference from the equilibrium experiment). This strategy is reflected in the politician's reputation. 

Suppose $\pi^{**}=\pi^*$, and the low-ability politician is obedient. Compare the reputations between the transparent case and the nontransparent case (\cref{eq basic reputation} and \ref{eq TP A reputations}):
\begin{equation}\label{eq reputation nontransparent preferences}
    \tau^{**}(a,A)\geq\tau^*(a,A)\text{ and }\tau^{**}(b,B)<\tau^*(b,B).
\end{equation}
We can see that the nontransparency of the lobbyist's preference increases the reputation for lobbyist A's preferred action but decreases the reputation for lobbyist A's disliked action. It also relaxes the obedience constraint: if $\pi^{**}=\pi^*$ and $\mu^{**}_{\Tilde{a}}=\mu^*_{\Tilde{a}}$,
\begin{equation}\label{eq slacked constrainst TP}
    2\mu^{**}_{\Tilde{a}}-1=\theta[(1-\mu^{*}_{\Tilde{a}})\tau^{*}(b,B)-\mu^{*}_{\Tilde{a}}\tau^{*}(a,A)]>\theta[(1-\mu^{**}_{\Tilde{a}})\tau^{**}(b,B)-\mu^{**}_{\Tilde{a}}\tau^{**}(a,A)].
\end{equation}
Therefore, lobbyist A can set $\mu^{**}_{\Tilde{a}}<\mu^*_{\Tilde{a}}$ to satisfy the obedience constraint (\cref{eq TP A equality}), which implies that $\pi^{**}$ is less informative than $\pi^*$.

The next proposition examines the effect of the politician's reputation concern when the lobbyist's preference is nontransparent.

\begin{prop}\label{prop harmful}
    Suppose that the lobbyist's preference is unknown to the public. If $\alpha\in(1-\mu_0-(1-2\mu_0)/(1-\tau),1-\mu_0)$, then:
    \begin{enumerate}[(i)]
        \item For any $\theta>0$, $\pi^{**}$ is unique and informative, and $\rho^{**}$ is essentially uninformative. In particular, $\mu^{**}_{\Tilde{a}}<1/2$ conditional on $\pi^{**}$.
        \item An increase in $\theta$ makes $\pi^{**}$ less informative and decreases the public's welfare.
    \end{enumerate}
\end{prop}

The proposition shows that when the lobbyist's preference is unknown to the public, the politician's reputation concern may induce the lobbyist to provide less information and harm the public's welfare. The result relies on the uncertainty of the lobbyist's bias. In particular, the probability that he is biased toward $a$ needs to be intermediate.

The proof can be sketched as follows. \cref{lem experiments nontransparent preferences} shows that if $\alpha$ is large enough, lobbyist A provides information in equilibrium, and \cref{eq TP A equality}, \ref{eq TP A posteriors}, and \ref{eq TP A reputations} characterize his equilibrium experiment. If $\alpha$ is also small enough, \cref{eq TP A reputations} shows $\tau^{**}(a,A)>\tau^{**}(b,B)$, and \cref{eq TP A equality} implies $\mu^{**}_{\Tilde{a}}<1/2$. In words, given $\Tilde{a}$ and $\pi^{**}$, action $a$ has lower decision quality than $b$, but $a$ is still taken by the low-ability politician because she seeks to enhance her reputation. If $\theta$ increases, the politician is more inclined to take $a$, and the obedience constraint is easier to satisfy. Therefore, lobbyist A can decrease $\mu^{**}_{\Tilde{a}}$ to make the low-ability politician obedient. This proves that an increase in $\theta$ makes $\pi^{**}$ less informative.

\cref{prop harmful} shows how the nontransparency of the lobbyist's preference changes the effect of the politician's reputation concern. When the lobbyist's preference is nontransparent, the public does not know the lobbyist's disliked action. They can only perceive one action as lobbyist-disliked, and this perceived disliked action may differ from the actual disliked action. In \cref{prop harmful}, because it is lobbyist A that provides information in equilibrium, the actual disliked action is $b$; but because the prior of the lobbyist being lobbyist A is small enough, the perceived disliked action is $a$. Therefore, when the politician's reputation concern incentivizes her to take the perceived disliked action, the lobbyist can provide less information to persuade the politician.\medskip

\noindent\textbf{Remark.} Both \cref{MR} and this section assume that the public observes the decision consequence. Nevertheless, because the logic behind \cref{prop LG comparative} and \cref{prop harmful} does not rely on the consequence transparency, the economic messages conveyed in these two sections will remain the same if the decision consequence is nontransparent.

\section{Nontransparency of Decision Consequences}\label{TC}
This section studies the nontransparency of decision consequences: how it affects information provision and interacts with the transparency of the lobbyist's preference.

Assume that decision consequences are nontransparent so that the public only observes the lobbyist's preference and the politician's action. In this setup, the politician's reputation is denoted by $\tau(a)$ and $\tau(b)$. Further, it can be decomposed as
\[\tau(a)=w(a)\tau(a,A)\text{ and }\tau(b)=w(b)\tau(b,B),\]
in which $w(x)$ represents the public's belief that action $x$ is correct, and $\tau(x,\omega)$ denotes the reputation from taking $x$ in state $\omega$. As usual, $\tau(a,B)=\tau(b,A)=0$.

Denote lobbyist A's equilibrium experiment as $\pi^\dagger$ and lobbyist B's equilibrium experiment as $\rho^\dagger$. Following \cref{lem no randomization} and \cref{coro obedience}, the low-ability politician takes $a$ after receiving $\Tilde{a}$ and $b$ after receiving $\Tilde{b}$, and to maximize the probability of sending $\Tilde{a}$, lobbyist A sets $\pi^\dagger(\Tilde{a}|A)=1$ and $\pi^\dagger(\Tilde{a}|B)$ such that the obedience constraint is binding. Thus, the following equations characterize his equilibrium experiment:
\begin{equation}\label{eq TC posterior}
    \text{(Posteriors)}\quad\mu^\dagger_{\Tilde{a}}=\frac{\mu_0}{\mu_0+(1-\mu_0)\pi^\dagger(\Tilde{a}|B)},\ \mu^\dagger_{\Tilde{b}}=0.
\end{equation}
\begin{equation}\label{eq TC reputation}
    \text{(Reputations)}\quad\tau^\dagger(a,A)=\tau,\  \tau^\dagger(b,B)=\frac{\tau}{\tau+(1-\tau)\pi^\dagger(\Tilde{b}|B)}.
\end{equation}
\begin{equation}\label{eq TC w}
\begin{aligned}
    \text{(The public's }&\text{belief about decision consequences)}\\
    &w^\dagger(a)=\frac{\mu_0}{\mu_0+(1-\mu_0)(1-\tau)\pi^\dagger(\Tilde{a}|B)},\  w^\dagger(b)=1.
\end{aligned}
\end{equation}
\begin{equation}\label{eq TC equality}
    \text{(The binding obedience constraint)}\quad 2\mu^\dagger_{\Tilde{a}}-1=\theta[w^\dagger(b)\tau^\dagger(b,B)-w^\dagger(a)\tau^\dagger(a,A)].
\end{equation}
We can see that they share similar structures to those in the baseline model. In particular, if ignoring the difference in the experiment, the reputations from taking correct actions are the same between the nontransparent and transparent cases (see \cref{eq basic reputation} and \cref{eq TC reputation}). Further, to evaluate the reputational payoffs, the politician discounts these reputations in both \cref{eq basic equality} and \cref{eq TC equality}.

The major difference arises in how these reputations are discounted. When the public observes the decision consequence, and the politician takes an incorrect action, the politician is revealed as having a low ability and loses all the reputation. Therefore, the reputational payoff from an action involves the posterior of the state, which gives rise to $\theta\mu_{\Tilde{a}}\tau(a,A)$ for $a$ and $\theta(1-\mu_{\Tilde{a}})\tau(b,B)$ for $b$. However, when the public only forms a belief about the decision consequence, the politician can earn a positive reputation without being correct. The reputational payoff from an action only involves the public's belief about decision consequences, which gives rise to $\theta w(a)\tau(a,A)$ for $a$ and $\theta w(b)\tau(b,B)$ for $b$. Unlike the transparent case, the public's belief about decision consequences depends on both the lobbyist's experiment and the politician's strategy. For example, because the low-ability politician knows the true state when she takes $b$, the politician who takes $b$ is deemed correct by the public, i.e., $w^\dagger(b)=1$.

Lobbyist B's equilibrium experiment, denoted by $\rho^\dagger$, can be characterized similarly:
\begin{equation}\label{eq LB TC posterior}
    \text{(Posteriors)}\quad\mu^\dagger_{\Tilde{a}}=1,\ \mu^\dagger_{\Tilde{b}}=\frac{\mu_0\rho^\dagger(\tilde{b}|A)}{\mu_0\rho^\dagger(\tilde{b}|A)+1-\mu_0}.
\end{equation}
\begin{equation}\label{eq LB TC reputation}
    \text{(Reputations)}\quad\tau^\dagger(a,A)=\frac{\tau}{\tau+(1-\tau)\rho^\dagger(\Tilde{a}|A)},\  \tau^\dagger(b,B)=\tau.
\end{equation}
\begin{equation}\label{eq LB TC w}
\begin{aligned}
    \text{(The public's }&\text{belief about decision consequences)}\\
    &w^\dagger(a)=1,\  w^\dagger(b)=\frac{1-\mu_0}{1-\mu_0+\mu_0(1-\tau)\rho^\dagger(\tilde{b}|A)}.
\end{aligned}
\end{equation}
\begin{equation}\label{eq LB TC equality}
    \text{(The binding obedience constraint)}\quad 2\mu^\dagger_{\Tilde{b}}-1=\theta[w^\dagger(b)\tau^\dagger(b,B)-w^\dagger(a)\tau^\dagger(a,A)].
\end{equation}
Because $\mu_0<1/2$, these equations may not admit a solution with $\mu^\dagger_{\Tilde{b}}\in(0,\mu_0)$. Let $\underline{\theta}^\dagger$ denote the value of $\theta>0$ such that \cref{eq LB TC equality} holds with $\rho^\dagger(\Tilde{a}|A)=0$ and $\mu^\dagger_{\Tilde{b}}=\mu_0$. This positive threshold exists for any $\mu_0$ and $\tau$.

The following lemma summarizes the result.

\begin{lem}\label{lem experiments TC}
    Suppose that decision consequences are nontransparent. 
    \begin{enumerate}[(i)]
        \item Lobbyist A's equilibrium experiment $\pi^\dagger$ is unique and determined by \cref{eq TC posterior}, \ref{eq TC reputation}, \ref{eq TC w} and \ref{eq TC equality}. In particular, $\mu^\dagger_{\Tilde{a}}>1/2$.
        \item Lobbyist B's equilibrium experiment $\rho^\dagger$ is informative if $\theta>\underline{\theta}^\dagger$, and essentially uninformative otherwise. When $\rho^\dagger$ must be informative, it is unique and determined by \cref{eq LB TC posterior}, \ref{eq LB TC reputation}, \ref{eq LB TC w} and \ref{eq LB TC equality}.
    \end{enumerate}
\end{lem} 

The next proposition shows the effect of revealing the decision consequence by comparing $\pi^*$ with $\pi^\dagger$ and $\rho^*$ with $\rho^\dagger$.

\begin{prop}\label{prop TC}
When the lobbyist's preference is known to the public, the transparency of decision consequences, compared with the nontransparency, decreases information provision and the public's welfare: $\pi^*<\pi^\dagger$ and $\rho^*\leq\rho^\dagger$.
\end{prop}

The proposition shows that revealing the decision consequence is harmful to the public. When the public observes the decision consequence, the politician has to be correct to earn a positive reputation. This requirement moderates the reputational incentive to take the lobbyist’s disliked action. As a result, the lobbyist can persuade the politician with less information provision when decision consequences are transparent.

Take the comparison between $\pi^*$ and $\pi^\dagger$ as an example. The intuition can be demonstrated by comparing the differences in reputational payoffs. Conditional on $\Tilde{a}$,
the difference in reputational payoffs between $b$ and $a$ is
\begin{equation}\label{eq reputation diff w/o TC}
    \theta[w(b)\tau(b,B)-w(a)\tau(a,A)] \text{ without transparent consequences},
 \end{equation}
but it is 
\begin{equation}\label{eq reputation diff w TC}
    \theta[(1-\mu_{\Tilde{a}})\tau(b,B)-\mu_{\tilde{a}}\tau(a,A)] \text{ with transparent consequences}.
\end{equation}
Other things equal, if the transparency of decision consequences indeed moderates the reputational incentive to take $b$, \eqref{eq reputation diff w TC} will be smaller than \eqref{eq reputation diff w/o TC}.

Suppose $\pi^*=\pi^\dagger$, and the low-ability politician is obedient. Because $\mu^*_{\Tilde{a}}=\mu^\dagger_{\Tilde{a}}>1/2$ and $w^\dagger(b)=1\geq w^\dagger(a)$, we know
\begin{equation}\label{eq TC comparing w}
    \frac{1-\mu^*_{\Tilde{a}}}{w^\dagger(b)}<\frac{\mu^*_{\Tilde{a}}}{w^\dagger(a)}.
\end{equation}
This inequality implies that the difference in reputational payoffs is smaller with transparent consequences, i.e.,
\begin{equation}\label{eq TC comparing reputation loss}
    \theta[(1-\mu^*_{\Tilde{a}})\tau^*(b,B)-\mu^*_{\Tilde{a}}\tau^*(a,A)]<\theta[w^\dagger(b)\tau^\dagger(b,B)-w^\dagger(a)\tau^\dagger(a,A)],
\end{equation}
and the transparency of decision consequences relaxes the obedience constraint: if $\pi^{*}=\pi^\dagger$ and $\mu^{*}_{\Tilde{a}}=\mu^\dagger_{\Tilde{a}}$,
\begin{equation}\label{eq TC-C slack constraint}
    2\mu^{*}_{\Tilde{a}}-1=\theta[w^{\dagger}(b)\tau^\dagger(b,B)-w^\dagger(a)\tau^\dagger(a,A)]>\theta[(1-\mu^{*}_{\Tilde{a}})\tau^*(b,B)-\mu^{*}_{\Tilde{a}}\tau^*(a,A)].
\end{equation}
Therefore, lobbyist A can set $\mu^*_{\Tilde{a}}<\mu^\dagger_{\Tilde{a}}$ to satisfy the obedience constraint (\cref{eq basic equality}), showing that $\pi^*$ is less informative than $\pi^\dagger$.

\cref{prop TC} starkly contrasts the finding in previous studies. The proposition shows that the transparency of decision consequences is harmful to the public, whereas \citet{prat2005wrong}, \citet{fox2012costly}, and \citet{FU201415} show the opposite possibility. To see the driving force of the difference, fix lobbyist A's experiment as $\pi^*$. When the decision consequence is transparent, the low-ability politician takes $a$ after receiving $\Tilde{a}$. This action is optimal for the public's welfare because $\mu^*_{\Tilde{a}}>1/2$. When the decision consequence is nontransparent, however, the low-ability politician may take $b$ after receiving $\Tilde{a}$ as $\pi^*$ does not satisfy the obedience constraint. This action is suboptimal for the public's welfare. Therefore, when information provision is exogenous, the transparency of decision consequences benefits the public, consistent with the finding in the literature. The example highlights endogenous information provision as the key feature driving the novel result about the consequence transparency.

The transparency of decision consequences may also be beneficial when the lobbyist's preference is unknown to the public. In the following, consider the setup with nontransparent preferences and nontransparent consequences, where the public only observes the politician's action. I assume:
\begin{as}\label{as TC}
$2\mu_0-1<\theta[\tau(1-\mu_0)/(1-\tau\mu_0)-1].$
\end{as}

Denote the equilibrium experiments of lobbyist A and lobbyist B with $\pi^\ddagger$ and $\rho^\ddagger$. \cref{as TC} guarantees that for any $\alpha\in(0,1)$, $\pi^\ddagger$ is unique and informative, and $\rho^\ddagger$ is essentially uninformative (which is formally shown in the proof of \cref{prop TC 2}). In particular, $\pi^\ddagger$ is characterized by the following equations:
\begin{equation}\label{eq TC 2 posteriors}
   \text{(Posteriors)}\quad\mu^\ddagger_{a}=\frac{\mu_0}{\mu_0+(1-\mu_0)\pi^\ddagger(\Tilde{a}|B)},\  \mu^\ddagger_{\Tilde{b}}=0.
\end{equation}
\begin{equation}\label{eq TC 2 reputations}
    \text{(Reputations)}\ \tau^\ddagger(a,A)=\frac{\tau}{\tau+(1-\tau)\alpha}, \tau^\ddagger(b,B)=\frac{\tau}{\tau+(1-\tau)[\alpha\pi^\ddagger(\Tilde{b}|B)+(1-\alpha)]}.
\end{equation}
\begin{equation}\label{eq TC 2 w}
\begin{aligned}
\text{(The public's }&\text{belief about decision consequences)}\\ 
&w^\ddagger(a)=\frac{\mu_0[\tau+(1-\tau)\alpha]}{\mu_0[\tau+(1-\tau)\alpha]+(1-\mu_0)(1-\tau)\alpha\pi^\ddagger(\Tilde{a}|B)}, \text{ and }\\
    &w^\ddagger(b)=\frac{(1-\mu_0)[\tau+(1-\tau)(\alpha\pi^\ddagger(\Tilde{b}|B)+1-\alpha)]}{(1-\mu_0)[\tau+(1-\tau)(\alpha\pi^\ddagger(\Tilde{b}|B)+1-\alpha)]+\mu_0(1-\tau)(1-\alpha)}.
\end{aligned}
\end{equation}
\begin{equation}\label{eq TC 2 equality}
    \text{(The binding obedience constraint)}\quad 2\mu^\ddagger_{\Tilde{a}}-1=\theta[w^\ddagger(b)\tau^\ddagger(b,B)-w^\ddagger(a)\tau^\ddagger(a,A)].
\end{equation}
The last equation implies that the low-ability politician is indifferent between $a$ and $b$ upon $\pi^{\ddagger}$ and $\mu^{\ddagger}_{\Tilde{a}}$. Further, it is optimal for her to take $a$ after receiving $\Tilde{a}$ from lobbyist A and take $b$ otherwise (the politician always knows the lobbyist's preference from the equilibrium experiment). This strategy is reflected in the politician's reputation and the public's belief about decision consequences.

Recall that $\pi^{**}$ and $\rho^{**}$ denote the equilibrium experiments when decision consequences are transparent, and the lobbyist's preference is unknown to the public. According to \cref{lem experiments nontransparent preferences} and \cref{as TC}, $\pi^{**}$ is unique and informative, and $\rho^{**}$ is essentially uninformative. The following proposition compares $\pi^{**}$ with $\pi^{\ddagger}$ and $\rho^{**}$ with $\rho^{\ddagger}$. For the comparison, I take essentially uninformative experiments as uninformative.

\begin{prop}\label{prop TC 2}
Suppose that the lobbyist's preference is unknown to the public, and \cref{as TC} holds. There exists $\underline{\alpha}>0$ such that if $\alpha\in(0,\underline{\alpha})$, the transparency of decision consequences, compared with the nontransparency, increases information provision and the public's welfare: $\pi^{**}>\pi^\ddagger$ and $\rho^{**}=\rho^\ddagger$.
\end{prop}

In contrast to \cref{prop TC}, the transparency of decision consequences may benefit the public when the lobbyist's preference is not publicly known. The difference arises because the politician's reputation concern may incentivize her to take the lobbyist's preferred action. In this case, by revealing the decision consequence, we can moderate the politician's reputational incentive and induce the lobbyist to provide more information to persuade the politician.

Formally, if $\alpha$ is sufficiently small, $\tau^\ddagger(a,A)>\tau^\ddagger(b,B)$ in \cref{eq TC 2 reputations} and $w^\ddagger(a)\approx 1>w^\ddagger(b)$ in \cref{eq TC 2 w}. Then, because the politician has a reputational incentive to take $a$, lobbyist A can set $\mu^\ddagger_{\Tilde{a}}<1/2$ in \cref{eq TC 2 equality}. In this case, the transparency of decision consequences can enhance information provision.

The result can be seen by examining how the difference in reputational payoffs between $b$ and $a$ changes with the consequence transparency. We can refer to \eqref{eq reputation diff w/o TC} and \eqref{eq reputation diff w TC} for these differences. Other things equal, \eqref{eq reputation diff w TC} will be larger than \eqref{eq reputation diff w/o TC} if the transparency of decision consequences indeed moderates the politician's incentive to take $a$. 

Suppose $\pi^{**}=\pi^\ddagger$, and the low-ability politician is obedient. Because $\mu^{**}_{\Tilde{a}}=\mu^\ddagger_{\Tilde{a}}<1/2$ and $w^\ddagger(a)>w^\ddagger(b)$, we know
\begin{equation}\label{eq TC 2 comparing w}
    \frac{1-\mu^{**}_{\Tilde{a}}}{w^\ddagger(b)}>\frac{\mu^{**}_{\Tilde{a}}}{w^\ddagger(a)}.
\end{equation}
This inequality implies that the difference in reputational payoffs is larger with transparent consequences, i.e., 
\begin{equation}\label{eq TC 2 change in reputational payoffs}
    \theta[(1-\mu^{**}_{\Tilde{a}})\tau^{**}(b,B)-\mu^{**}_{\Tilde{a}}\tau^{**}(a,A)]>\theta[w^\ddagger(b)\tau^\ddagger(b,B)-w^\ddagger(a)\tau^\ddagger(a,A)],
\end{equation}
and the transparency of decision consequences makes the obedience constraint more stringent: if $\pi^{**}=\pi^\ddagger$ and $\mu^{**}_{\Tilde{a}}=\mu^\ddagger_{\Tilde{a}}$,
\begin{equation}\label{eq C-TC stringent}
2\mu^{**}_{\Tilde{a}}-1=\theta[w^\ddagger(b)\tau^{\ddagger}(b,B)-w^\ddagger(a)\tau^{\ddagger}(a,A)]<\theta[(1-\mu^{**}_{\Tilde{a}})\tau^{**}(b,B)-\mu^{**}_{\Tilde{a}}\tau^{**}(a,A)].
\end{equation}
Hence, to satisfy the obedience constraint (\cref{eq TP A equality}), lobbyist A must set $\mu^{**}_{\Tilde{a}}>\mu^\ddagger_{\Tilde{a}}$, showing that $\pi^{**}$ is more informative than $\pi^\ddagger$.

\cref{prop TC} and \cref{prop TC 2} demonstrate the conflict between the transparency of the lobbyist's preference and the transparency of decision consequences. In particular, the transparency of the lobbyist's preference distorts the effect of revealing the decision consequence. In this sense, the finding cautions against disclosing lobbyists' preferences. However, because \cref{prop LG comparative} can be adapted to the setup with nontransparent consequences, we can conclude from \cref{prop LG comparative} and \cref{prop TC} that the public's welfare is maximized by revealing the lobbyist's preference and hiding the decision consequence.

\section{Discussion}\label{D}
Beyond the analysis presented above, many other transparency forms exist to consider. The action taken by the politician may be confidential due to national security concerns. The lobbyist's recommendation may be transparent when the lobbyist reveals his recommendation to the media. This section examines these transparency forms. The formal results are shown in Appendix B.\footnote{This appendix and Appendix C (mentioned in \cref{C}) are both placed on my personal website.} \bigskip

\noindent\textbf{Nontransparency of the Politician's Action.} Consider the setup where the public does not observe the politician's action but observes the decision consequence. In this setup, the decision consequence determines the politician's reputation and maximizes it when the politician's action is correct. Thus, the politician's reputation concern is aligned with her concern for decision quality, leading her to behave as if she has no reputation concern. In other words, the nontransparency of the politician's action eliminates the influence of the politician's reputation concern.

When the lobbyist's preference is transparent, the politician's reputation concern can enhance information provision (\cref{prop beneficial}). It is harmful to eliminate this positive influence by hiding the politician's action. When the lobbyist's preference is nontransparent, the politician's reputation concern may impede information provision (\cref{prop harmful}). It is beneficial to eliminate this negative influence by hiding the politician's action. Consequently, the nontransparency of the politician's action may benefit the public only when the lobbyist's preference is nontransparent.

When the lobbyist's preference is transparent, the result contrasts with the finding in the literature. In particular, it shows that the nontransparency of the politician's action harms the public's welfare, whereas \citet{prat2005wrong} and \citet{fox2012costly} show the opposite possibility. The difference is again driven by endogenous information provision. If information is exogenous, the nontransparency of the politician's action benefits the public because it aligns the politician's incentive with the public's welfare. If information is endogenous, however, the nontransparency of the politician's action removes the politician's incentive to take the lobbyist's disliked action. This effect allows the lobbyist to persuade the politician with less information provision.\bigskip

\noindent\textbf{Transparency of the Lobbyist's Recommendation.} Suppose that the public observes what action is recommended. Then, the politician can enhance her reputation by contradicting the lobbyist's recommendation. When the lobbyist publicly recommends his preferred action, the politician's incentive to take the lobbyist's disliked action is stronger compared with the nontransparent case. Therefore, with transparent recommendations, the lobbyist must provide more information to persuade the politician. The transparency of the lobbyist's recommendation benefits the public by enhancing information provision.

Take lobbyist A as an example. In equilibrium, when he recommends action $a$ by sending $\Tilde{a}$, a low-ability politician obeys and takes $a$, but a high-ability politician may disobey and take $b$. Hence, conditional on $\tilde{a}$, the public perceives the politician as high-ability if she takes $b$. Formally,
\begin{equation}
    \tau(a,A,\Tilde{a})=\tau\text{ (note $\pi(\Tilde{a}|A)=1$) and } \tau(b,B,\tilde{a})=1,
\end{equation}
in which $\tau(x,\omega,s)$ denotes the reputation from taking action $x$ upon state $\omega$ and recommendation $s$. Note that $\tau(a,A,\Tilde{a})=\tau^*(a,A)$ and $\tau(b,B,\Tilde{a})>\tau^*(b,B)$ (see \cref{eq basic reputation}). If the lobbyist designs the same experiment as $\pi^*$ (see \cref{lem equilibrium}), the obedience constraint will be violated: given $\pi=\pi^*$ and $\mu_{\Tilde{a}}=\mu^*_{\Tilde{a}}$,
\begin{equation}
    2\mu_{\Tilde{a}}-1=\theta[(1-\mu^*_{\Tilde{a}})\tau^*(b,B)-\mu^*_{\Tilde{a}}\tau^*(a,A)]<\theta[(1-\mu_{\Tilde{a}})\tau(b,B,\Tilde{a})-\mu_{\Tilde{a}}\tau(a,A,\Tilde{a})].
\end{equation}
Therefore, lobbyist A must set $\mu_{\Tilde{a}}>\mu^*_{\Tilde{a}}$ to make the low-ability politician obedient and design an experiment more informative than $\pi^*$. Compared with the baseline model, where the lobbyist's recommendation is nontransparent, the lobbyist provides more information when the public observes the recommended action.\footnote{A related result about how a sender privately persuades voters can be found in \citet{CHAN2019}.}

\section{Conclusion}\label{C}
This article shows how the politician's reputation concern affects endogenous information provision and how the effect depends on the transparency design. The result emphasizes the transparency of the lobbyist's preference. When the lobbyist's preference is transparent, the politician's reputation concern induces biased lobbyists to provide more information. But when the lobbyist's preference is nontransparent, the politician's reputation concern may have the opposite effect.

Moreover, the comparison between transparent and nontransparent preferences shows that the transparency of the lobbyist's preference induces biased lobbyists to provide more information, helping the politician make a better policy. This result provides a rationale for adopting transparent lobbying registers.

The article also highlights a conflict between the conventional transparency of decision consequences and the new transparency of the lobbyist’s preference. In particular, when the lobbyist’s preference is transparent, the nontransparency of decision consequences is better than transparency.

Throughout the article, the politician is assumed to have perfect knowledge of her ability, representing one end of the spectrum. On the other end, the politician knows nothing about her ability. Between them lies a continuum of settings where the politician has an imperfect signal about her ability. Appendix C provides an analysis of these settings. It is shown that the politician's reputation concern can still induce the lobbyist to provide more information as long as she has a sufficiently precise signal about her ability.

\bibliography{ref0}

\appendix

\section{Proofs}\label{App1}

\subsection{Proof of Lemma \texorpdfstring{\ref{lem no randomization}}{\ref*{lem no randomization}}}
Consider the baseline model or any transparency setting in \cref{LG} and \cref{TC}. The public observes the politician's action but may not observe the decision consequence or the lobbyist's preference. Suppose that one type of lobbyist designs an experiment $\pi$ on the equilibrium path such that the low-ability politician randomizes between $a$ and $b$ after some recommendation.\footnote{I do not assume the binary recommendation space $\{\Tilde{a},\Tilde{b}\}$ in this proof.} Denote this recommendation as $\tilde{c}$ and the posterior conditional on $\tilde{c}$ and $\pi$ as $\mu_{\Tilde{c}}$. Denote as $p\in(0,1)$ the probability that the low-ability politician assigns to $a$ upon $\mu_{\tilde{c}}$. I intend to show another experiment that can improve the lobbyist's payoff.

Because the low-ability politician's action is optimal, the payoffs from $a$ and $b$ must be equal upon $\mu_{\tilde{c}}$, i.e., $\bfE^{\mu_{\tilde{c}}}[u^P(a,\omega)-u^P(b,\omega)]=0.$ When the public observes the decision consequence, the equation can be written as
\begin{equation}\label{eq proof randomization 1}
    2\mu_{\tilde{c}}-1+\theta[\mu_{\tilde{c}}\tau(a,A)-(1-\mu_{\tilde{c}})\tau(b,B)]=0.
\end{equation}
And when the public cannot observe the decision consequence, the equation can be written as
\begin{equation}\label{eq proof randomization 2}
    2\mu_{\tilde{c}}-1+\theta[\tau(a)-\tau(b)]=0,
\end{equation}
in which $\tau(x)$ denotes the reputation from taking action $x$. In both cases, the payoff difference increases in the posterior for fixed reputations. Hence, it is strictly optimal for the low-ability politician to take $a$ upon a posterior greater than $\mu_{\Tilde{c}}$ and to take $b$ upon a posterior smaller than $\mu_{\Tilde{c}}$. 

Without loss of generality, assume that $\pi$ induces three possible posteriors, $\mu_{\tilde{b}}<\mu_{\tilde{c}}<\mu_{\Tilde{a}}$, and the low-ability politician takes $a$ upon $\mu_{\tilde{a}}$, $b$ upon $\mu_{\Tilde{b}}$, and randomizes upon $\mu_{\tilde{c}}$. Define $q_{s}:=\mu_0\pi(s|A)+(1-\mu_0)\pi(s|B)$ for $s\in\{\Tilde{b},\Tilde{c},\Tilde{a}\}$. By assumption, $q_{\tilde{c}}>0$. If $\Tilde{a}$ or $\Tilde{b}$ is not defined, $q_{\Tilde{a}}$ or $q_{\Tilde{b}}$ is taken as $0$.

Suppose that the lobbyist who designs $\pi$ on the equilibrium path is lobbyist A (the case for lobbyist B can be proven similarly). There are two cases: (1) $\mu_{\Tilde{c}}<1$ and (2) $\mu_{\Tilde{c}}=1$. 

For the first case, I construct $\pi'$ as follows. If $q_{\tilde{c}}=1$, then let $\pi'$ be an experiment that induces two possible posteriors, $\mu'_{\tilde{b}}=0$ and $\mu'_{\tilde{c}}$, such that $\mu'_{\tilde{c}}>\mu_0$. If $q_{\tilde{c}}<1$, then let $\pi'$ be an experiment that induces three possible posteriors, $\mu'_{\Tilde{b}}$, $\mu'_{\tilde{c}}$, and $\mu'_{\tilde{a}}$, such that $\pi'$ differs from $\pi$ in $\mu'_{\tilde{c}}>\mu_{\tilde{c}}$, but $\mu'_{\Tilde{b}}=\mu_{\Tilde{b}}$ and $\mu'_{\Tilde{a}}=\mu_{\Tilde{a}}$. Let $q'_{\tilde{b}}, q'_{\tilde{c}}$ and $q'_{\tilde{a}}$ denote the probability of each recommendation being sent under $\pi'$. \citet{kamenica2011bayesian} has proven that
\begin{equation}\label{eq proof randomization martingale}
    \sum_{i\in\{\Tilde{b},{\tilde{c}},\Tilde{a}\}}q_i\mu_i=\mu_0,\text{ and } \sum_{i\in\{\Tilde{b},{\tilde{c}},\Tilde{a}\}}q'_i\mu'_i=\mu_0.
\end{equation}

Let lobbyist A deviate to $\pi'$. The reputation does not change because the public does not observe the experiment. If the recommendation is $\Tilde{a}$ or $\Tilde{b}$, the optimal action of the low-ability politician does not change, either. However, if the recommendation is $\Tilde{c}$, the low-ability politician optimally takes $a$ because $\mu'_{\tilde{c}}>\mu_{\tilde{c}}$. Given that the politician is low-ability, lobbyist A's payoff from $\pi$ is
$$q_{\Tilde{a}}+q_{\Tilde{c}}p,$$
and his payoff from $\pi'$ is 
$$q'_{\Tilde{a}}+q'_{\Tilde{c}}.$$
Because of \cref{eq proof randomization martingale} and $\sum_s q_s=\sum_s q'_s=1$, $q'_{s}\to q_s$ as $\mu'_{\Tilde{c}}\to\mu_{\tilde{c}}$ for any $s\in\{\Tilde{b},\Tilde{c},\Tilde{a}\}$. In the limit, lobbyist A can improve his payoff by
\[q'_{\tilde{c}}-q_{\tilde{c}} p+(q'_{\Tilde{a}}-q_{\Tilde{a}})\to q_{\tilde{c}}(1-p)>0.\]
Hence, there exists another experiment that is a profitable deviation for lobbyist A.

The proof is complete if $\mu_{\Tilde{c}}\not=1$. Instead, suppose $\mu_{\Tilde{c}}=1$. According to \cref{eq proof randomization 1}, this case cannot happen when the public observes the decision consequence. In the following, assume that the public does not observe the decision consequence. 

By $\pi$, lobbyist A sends two possible recommendations, $\Tilde{b}$ and $\Tilde{c}$. Given that the politician is low-ability, the lobbyist's payoff from $\pi$ is $q_{\Tilde{c}}p$. The optimality of $\pi$ requires $\mu_{\Tilde{b}}=0$ such that $q_{\Tilde{c}}$ can be maximized. Hence, $\pi$ is full information disclosure, and $q_{\Tilde{b}}=1-\mu_0$ and $q_{\Tilde{c}}=\mu_0$ in \cref{eq proof randomization martingale}. Then, because the high-ability politician takes $a$ in $A$, but the low-ability politician takes $a$ only with probability $p$ in $A$, we conclude $\tau(a)>\tau(b)$ if the public observes the lobbyist's preference:
$$\tau(a)=\frac{\tau\mu_0}{\tau\mu_0+(1-\tau)\mu_0 p}>\tau(b)=\frac{\tau(1-\mu_0)}{\tau(1-\mu_0)+(1-\tau)[1-\mu_0+\mu_0(1-p)]}.$$
On the other hand, if the public does not observe the lobbyist's preference, it is optimal for lobbyist B to reveal no information (the low-ability politician will always take $b$ for him because $\mu_0<\mu_{\Tilde{c}}=1$). Again, we have $\tau(a)>\tau(b)$:
\begin{equation}
    \tau(a)=\frac{\tau\mu_0}{\tau\mu_0+(1-\tau)\alpha\mu_0 p}>\tau(b)=\frac{\tau(1-\mu_0)}{\tau(1-\mu_0)+(1-\tau)(1-\alpha\mu_0 p)}.
\end{equation}
However, then we have $\bfE^{\mu_{\Tilde{c}=1}}[u^P(a,\omega)-u^P(b,\omega)]=1+\theta[\tau(a)-\tau(b)]>0$, which contradicts with \cref{eq proof randomization 2}. As a result, $\mu_{\Tilde{c}}\not=1$, and the proof is complete. \qed

\subsection{Proof of Corollary \texorpdfstring{\ref{coro obedience}}{\ref*{coro obedience}}}\label{proof coro obedience}
The first step of the proof is to rule out two cases: (1) the low-ability politician always takes the lobbyist's disliked action, and (2) the low-ability politician takes $a$ after receiving $\Tilde{b}$ and $b$ after receiving $\Tilde{a}$. The first case does not happen in equilibrium because, given a low-ability politician, full information disclosure guarantees a positive payoff for any lobbyist. It is suboptimal for any lobbyist to let the low-ability politician always take his disliked action. The second case does not happen in equilibrium because $\mu_{\tilde{b}}\leq\mu_{\tilde{a}}$, and the payoff difference between $a$ and $b$ increases in the posterior for fixed reputations (see \cref{eq proof randomization 1} and \ref{eq proof randomization 2} for the details).

Then, according to \cref{lem no randomization}, only the two cases mentioned in the corollary remain. When the equilibrium experiment is uninformative, it only sends one recommendation. The politician, who takes a pure action (\cref{lem no randomization}), must take the lobbyist's preferred action. When the equilibrium experiment cannot be uninformative, suppose that the low-ability politician always takes the lobbyist's preferred action. Then, because the payoff difference between $a$ and $b$ is monotone for fixed reputations, this experiment can be replaced with an uninformative experiment such that the low-ability politician does not change her behavior. This is a contradiction. Hence, when the equilibrium experiment is not essentially uninformative, the low-ability politician takes $a$ after receiving $\Tilde{a}$ and $b$ after receiving $\Tilde{b}$.\qed

\subsection{Proof of Lemma \texorpdfstring{\ref{lem equilibrium}}{\ref*{lem equilibrium}}}\label{proof lem equilibrium}
For the lemma, I only show the unique existence of the experiments characterized by \cref{eq basic posterior}, \ref{eq basic reputation}, and \ref{eq basic equality} and characterized by \cref{eq LB basic posterior}, \ref{eq LB basic reputation}, and \ref{eq LB basic equality}. The argument for why these equations characterize $\pi^*$ and $\rho^*$ is standard and mentioned in the main text.

For $\pi^*$, substitute \cref{eq basic posterior} and \ref{eq basic reputation} into \cref{eq basic equality}. The binding obedience constraint can be written as
\begin{equation}
    2\mu^*_{\Tilde{a}}-1=\theta\left[(1-\mu^*_{\Tilde{a}})\cdot\frac{\tau}{\tau+(1-\tau)(1-\mu_0(1-\mu^*_{\Tilde{a}})/(1-\mu_0)\mu^*_{\Tilde{a}})}-\mu^*_{\Tilde{a}}\tau\right].
\end{equation}
Define $F(\mu,\theta)$ as the difference between the left-hand side and the right-hand side, and $F(\mu^*_{\Tilde{a}},\theta)=0$. Note that (1) $F(\cdot,\theta)$ is continuous and monotone in $\mu$, (2) $F(\mu_0,\theta)<0$, and (3) $F(1,\theta)>0$. Hence, for any $\theta$, there exists a unique $\mu^*_{\Tilde{a}}\in(\mu_0,1)$ such that $F(\mu^*_{\Tilde{a}},\theta)=0$. In particular, $\mu^*_{\Tilde{a}}>1/2$ because $F(1/2,\theta)<0$.

For $\rho^*$, substitute \cref{eq LB basic posterior} and \ref{eq LB basic reputation} into \cref{eq LB basic equality}. The binding obedience constraint can be written as
\begin{equation}
    2\mu^*_{\Tilde{b}}-1=\theta\left[(1-\mu^*_{\Tilde{b}})\tau-\mu^*_{\Tilde{b}}\cdot\frac{\tau}{\tau+(1-\tau)(1-\mu^*_{\Tilde{b}}(1-\mu_0)/\mu_0(1-\mu^*_{\Tilde{b}}))}\right].
\end{equation}
Define $L(\mu,\theta)$ as the difference between the left-hand side and the right-hand side, and $L(\mu^*_{\Tilde{b}},\theta)=0$. Note that (1) $L(\cdot,\theta)$ is continuous and monotone in $\mu$, and (2) $L(0,\theta)<0$. Hence, for fixed $\theta$, there exists $\mu^*_{\Tilde{b}}\in(0,\mu_0)$ such that $L(\mu^*_{\Tilde{b}},\theta)=0$ if and only if $$L(\mu_0,\theta)=2\mu_0-1-\theta[(1-\mu_0)\tau-\mu_0]>0.$$ This is impossible if $\tau\geq\mu_0/(1-\mu_0)$ and $\theta\geq 0$. In contrast, if $\tau<\mu_0/(1-\mu_0)$, then $\underline{\theta}>0$ with $L(\mu_0,\underline{\theta})=0$ exists, and $L(\mu_0,\theta)>0$ for $\theta>\underline{\theta}$. Consequently, if $\tau<\mu_0/(1-\mu_0)$ and $\theta>\underline{\theta}$, \cref{eq LB basic posterior}, \ref{eq LB basic reputation}, and \ref{eq LB basic equality} admit an informative $\rho^*$. Otherwise, $\rho^*$ can be uninformative. \qed

\subsection{Proof of Proposition \texorpdfstring{\ref{prop beneficial}}{\ref*{prop beneficial}}}
For $\pi^*$, stick to the definition of $F(\mu,\theta)$ in \cref{proof lem equilibrium}. Note that $\mu^*_{\Tilde{a}}$ is the solution to $F(\mu,\theta)=0$. We have 
\begin{equation}
\begin{aligned}
    \frac{\pd\mu^*_{\Tilde{a}}(\theta)}{\pd\theta}=&-\frac{\pd F(\mu^*_{\Tilde{a}},\theta)/\pd\theta}{\pd F(\mu^*_{\Tilde{a}},\theta)/\pd\mu^*_{\Tilde{a}}}\\
    =&\frac{1}{\pd F(\mu^*_{\Tilde{a}},\theta)/\pd\mu^*_{\Tilde{a}}}\left[(1-\mu^*_{\Tilde{a}})\cdot\frac{\tau}{\tau+(1-\tau)(1-\mu_0(1-\mu^*_{\Tilde{a}})/(1-\mu_0)\mu^*_{\Tilde{a}})}-\mu^*_{\Tilde{a}}\tau\right]\\
    =&\frac{(2\mu^*_{\Tilde{a}}-1)/\theta}{\pd F(\mu^*_{\Tilde{a}},\theta)/\pd\mu^*_{\Tilde{a}}}>0.
\end{aligned}
\end{equation}
The third equality holds because $F(\mu^*_{\tilde{a}},\theta)=0$. The derivative is positive because $\mu^*_{\Tilde{a}}>1/2$ and $F(\mu,\theta)$ is monotone in $\mu$. Given the common prior $\mu_0$, a distribution of posteriors on $\mu^*_1$ and $0$ strictly dominates another distribution on $\mu^*_2$ and $0$ in the Blackwell order if and only if $\mu^*_1>\mu^*_2$. Therefore, $\pi^*$ is increasingly informative as $\theta$ increases. In addition, the public's welfare increases in $\theta$ because the obedient politician with a larger $\theta$ receives more information. 

The proof for $\rho^*$ is similar. In particular, $\pd\mu^*_{\Tilde{b}}(\theta)/\pd\theta\leq 0$ for $\rho^*$.\qed

\subsection{Proof of Lemma \texorpdfstring{\ref{lem experiments nontransparent preferences}}{\ref*{lem experiments nontransparent preferences}}}\label{proof lem experiments nontransparent preferences}
First, I prove that if $\pi^{**}$ is unique and informative, $\rho^{**}$ is essentially uninformative, and vice versa. Suppose that $\pi^{**}$ is unique and informative (the other way around is similar). \cref{coro obedience} says that the low-ability politician takes $a$ after receiving $\Tilde{a}$ and $b$ after receiving $\Tilde{b}$. Then, to maximize the probability of sending $\Tilde{a}$, lobbyist A makes the obedience constraint binding after sending $\Tilde{a}$ (\cref{eq TP A equality}):
    $2\mu^{**}_{\Tilde{a}}-1=\theta[(1-\mu^{**}_{\Tilde{a}})\tau^{**}(b,B)-\mu^{**}_{\Tilde{a}}\tau^{**}(a,A)].$
Notice that $\mu^{**}_{\Tilde{a}}>\mu_0$. In this case, lobbyist B can design an experiment, $\rho$, such that $\mu_{\Tilde{b}}\leq\mu_{\Tilde{a}}<\mu^{**}_{\Tilde{a}}$ and $
    2\mu_{\Tilde{a}}-1<\theta[(1-\mu_{\Tilde{a}})\tau^{**}(b,B)-\mu_{\Tilde{a}}\tau^{**}(a,A)]$.
In particular, $\rho$ can be designed as inducing $\mu_{\Tilde{b}}=\mu_{\Tilde{a}}=\mu_0$. This experiment induces the low-ability politician to take $b$ upon any recommendation, which results in an outcome that is optimal for lobbyist B. Hence, lobbyist B's equilibrium experiment $\rho^{**}$ is essentially uninformative.

Below, I derive the sufficient and necessary condition under which an equilibrium experiment is unique and informative.

Assume that $\pi^{**}$ is unique and informative. To maximize the probability of sending $\Tilde{a}$, lobbyist A must set $\pi^{**}(\Tilde{a}|A)=1$ and make the obedience constraint binding after sending $\Tilde{a}$. Hence, \cref{eq TP A posteriors} gives the posteriors, and \cref{eq TP A reputations} gives the reputations. Substitute these two equations into \cref{eq TP A equality}. The binding obedience constraint can be written as
\begin{equation}\label{eq proof LG equality}
    2\mu^{**}_{\Tilde{a}}-1=\theta\left[(1-\mu^{**}_{\Tilde{a}})\cdot\frac{\tau}{\tau+(1-\tau)(1-\alpha\mu_0(1-\mu^{**}_{\Tilde{a}})/(1-\mu_0)\mu^{**}_{\Tilde{a}})}-\mu^{**}_{\Tilde{a}}\cdot\frac{\tau}{\tau+(1-\tau)\alpha}\right].
\end{equation}
Define $G(\mu,\theta)$ as the difference between the left-hand side and the right-hand side, and $G(\mu^{**}_{\Tilde{a}},\theta)=0$. Because $G(\mu,\theta)$ is continuous and monotone in $\mu$, if $\mu^{**}_{\Tilde{a}}\in[\mu_0,1]$ exists, it must be unique. The conditions for the existence are $G(1,\theta)\geq 0$ and $G(\mu_0,\theta)\leq 0$. The former is automatically true, but the latter requires
\begin{equation}\label{eq proof TP conditions}
    2\mu_0-1\leq\theta\left[(1-\mu_0)\frac{\tau}{\tau+(1-\tau)(1-\alpha)}-\mu_0\cdot\frac{\tau}{\tau+(1-\tau)\alpha}\right].
\end{equation}
If \eqref{eq proof TP conditions} is binding or violated, lobbyist A can induce the low-ability politician always to take $a$ by designing an uninformative experiment. Then, $\pi^{**}$ is essentially uninformative. If \eqref{eq proof TP conditions} is strict, lobbyist A must provide information to persuade the low-ability politician to take $a$. The strict version of \eqref{eq proof TP conditions} guarantees that \cref{eq proof LG equality} has a solution with $\mu^{**}_{\Tilde{a}}>\mu_0$. Then, $\pi^{**}$ is unique and informative.

Now, assume that $\rho^{**}$ is unique and informative. Similar to the argument in the previous paragraph, this experiment is characterized by the following equations:
\begin{equation}\label{eq TP B posteriors}
    \text{(Posteriors)}\quad\mu^{**}_{\Tilde{a}}=1,\  \mu^{**}_{\Tilde{b}}=\frac{\mu_0\rho^{**}(\Tilde{b}|A)}{\mu_0\rho^{**}(\Tilde{b}|A)+(1-\mu_0)}.
\end{equation}
\begin{equation}\label{eq TP B reputations}
    \text{(Reputations)} \tau^{**}(a,A)=\frac{\tau}{\tau+(1-\tau)[\alpha+(1-\alpha)\rho^{**}(\Tilde{a}|A)]},\ \tau^{**}(b,B)=\frac{\tau}{\tau+(1-\tau)(1-\alpha)}.
\end{equation}
\begin{equation}\label{eq TP B equality}
    \text{(The binding obedience constraint)}\quad 2\mu^{**}_{\Tilde{b}}-1=\theta[(1-\mu^{**}_{\Tilde{b}})\tau^{**}(b,B)-\mu^{**}_{\Tilde{b}}\tau^{**}(a,A)].
\end{equation}
Substitute the first two equations into the last one. The binding obedience constraint can be written as
\begin{equation}\label{eq proof LG B equality}
    2\mu^{**}_{\Tilde{b}}-1=\theta\left[(1-\mu^{**}_{\Tilde{b}})\cdot\frac{\tau}{\tau+(1-\tau)(1-\alpha)}-\mu^{**}_{\Tilde{b}}\cdot\frac{\tau}{\tau+(1-\tau)(1-(1-\alpha)\frac{(1-\mu_0)\mu^{**}_{\Tilde{b}}}{\mu_0(1-\mu^{**}_{\Tilde{b}})})}\right].
\end{equation}
Define $B(\mu,\theta)$ as the difference between the left-hand side and the right-hand side, and $B(\mu^{**}_{\Tilde{b}},\theta)=0$. Because $B(\mu,\theta)$ is continuous and monotone in $\mu$, if $\mu^{**}_{\Tilde{b}}\in[0,\mu_0]$ exists, it must be unique. The conditions for the existence are $B(0,\theta)\leq 0$ and $B(\mu_0,\theta)\geq 0$. The former is automatically true, but the latter requires
\begin{equation}\label{eq proof TP B conditions}
    2\mu_0-1\geq\theta\left[(1-\mu_0)\frac{\tau}{\tau+(1-\tau)(1-\alpha)}-\mu_0\cdot\frac{\tau}{\tau+(1-\tau)\alpha}\right].
\end{equation}
If \eqref{eq proof TP B conditions} is binding or violated, lobbyist B can induce the low-ability politician always to take $b$ by designing an uninformative experiment. Then, $\rho^{**}$ is essentially uninformative. If \eqref{eq proof TP B conditions} is strict, lobbyist B must provide information to persuade the low-ability politician to take $b$. The strict version of \eqref{eq proof TP B conditions} guarantees that \cref{eq proof LG B equality} has a solution with $\mu^{**}_{\Tilde{b}}<\mu_0$. Then, $\rho^{**}$ is unique and informative. \qed

\subsection{Proof of Proposition \texorpdfstring{\ref{prop LG comparative}}{\ref*{prop LG comparative}}}
Consider the case in which $\pi^{**}$ is unique and informative, and $\rho^{**}$ is uninformative. Clearly, $\rho^{**}\leq\rho^{*}$. To compare $\pi^{*}$ and $\pi^{**}$, suppose that $\pi^{**}=\pi^*$, and the low-ability politician is obedient. We can compare the reputation between the transparent and nontransparent cases from \cref{eq basic reputation} and \cref{eq TP A reputations}. The result, according to \cref{eq reputation nontransparent preferences}, is that $\tau^*(a,A)\leq\tau^{**}(a,A)$, and $\tau^*(b,B)>\tau^{**}(b,B)$. Then, \cref{eq reputation nontransparent preferences} and the binding obedience constraint satisfied by $\pi^*$ (\cref{eq basic equality}) imply that if $\pi^{**}=\pi^*$, the obedience constraint is slacked as \cref{eq slacked constrainst TP}. 

Note that $G(\mu,\theta)$ defined in \cref{proof lem experiments nontransparent preferences} is monotone in $\mu$. \cref{eq slacked constrainst TP} shows $G(\mu^*_{\Tilde{a}},\theta)>0$. Because $G(\mu^{**}_{\Tilde{a}},\theta)=0$, we conclude $\mu^{**}_{\Tilde{a}}<\mu^{*}_{\Tilde{a}}$. This fact with $\mu^{**}_{\Tilde{b}}=\mu^{*}_{\Tilde{b}}=0$ shows $\pi^{**}<\pi^*$. As a result, $\pi^{**}<\pi^*$ and $\rho^{**}\leq\rho^{*}$. The nontransparency of the lobbyist's preference decreases the public's welfare because it decreases the information provision for the obedient politician.

When $\rho^{**}$ is unique and informative, and $\pi^{**}$ is uninformative, the result can be proven similarly. \qed

\subsection{Proof of Proposition \texorpdfstring{\ref{prop harmful}}{\ref*{prop harmful}}}
Given $\alpha>1-\mu_0-(1-2\mu_0)/(1-\tau)$, we have
\[2\mu_0-1<\theta[(1-\mu_0)\tau/(\tau+(1-\tau)(1-\alpha))-\mu_0\tau/(\tau+(1-\tau)\alpha)], \forall\theta\geq 0.\]
According to \cref{lem experiments nontransparent preferences}, $\pi^{**}$ is unique and informative, and $\rho^{**}$ is uninformative.

Stick to the definition of $G(\mu,\theta)$ in \cref{proof lem experiments nontransparent preferences}. Examine $\mu^{**}_{\Tilde{a}}$ by noting $G(\mu^{**}_{\Tilde{a}},\theta)=0$. To show $\mu^{**}_{\Tilde{a}}<1/2$, it suffices to show $G(1/2,\theta)>0$. Indeed,
\begin{equation}
    G\left(\frac{1}{2},\theta\right)=\frac{\theta\tau(1-\tau)}{2[\tau+(1-\tau)(1-\alpha\mu_0/(1-\mu_0))](\tau+(1-\tau)\alpha)}\left(1-\frac{1}{1-\mu_0}\alpha\right),
\end{equation}
which is positive because $\alpha<1-\mu_0$. Then, from $G(\mu^{**}_{\Tilde{a}},\theta)=0$ and $\mu^{**}_{\Tilde{a}}<1/2$, we can show that
\begin{align}
    \frac{\pd\mu^{**}_{\Tilde{a}}(\theta)}{\pd\theta}=&-\frac{\pd G(\mu^{**}_{\Tilde{a}},\theta)/\pd\theta}{\pd G(\mu^{**}_{\Tilde{a}},\theta)/\pd\mu^{**}_{\Tilde{a}}}\\
   =&\frac{1}{\pd G(\mu^{**}_{\Tilde{a}},\theta)/\pd\mu^{**}_{\Tilde{a}}}\left[\frac{(1-\mu^{**}_{\Tilde{a}})\tau}{\tau+(1-\tau)(1-\alpha\mu_0(1-\mu^{**}_{\Tilde{a}})/(1-\mu_0)\mu^{**}_{\Tilde{a}})}-\frac{\mu^{**}_{\Tilde{a}}\tau}{\tau+(1-\tau)\alpha}\right]\\
=&\frac{(2\mu^{**}_{\Tilde{a}}-1)/\theta}{\pd G(\mu^{**}_{\Tilde{a}},\theta)/\pd\mu^{**}_{\Tilde{a}}}<0.
\end{align}
The third equality holds because $G(\mu^{**}_{\Tilde{a}},\theta)=0$. The derivative is negative because $\mu^{**}_{\Tilde{a}}<1/2$ and $G(\mu,\theta)$ is monotone in $\mu$. As a result, $\pi^{**}$ is decreasingly information as $\theta$ increases. The public's welfare decreases in $\theta$ because the obedient politician with a larger $\theta$ receives less information. \qed

\subsection{Proof of Lemma \texorpdfstring{\ref{lem experiments TC}}{\ref*{lem experiments TC}}}\label{proof lem experiments TC}
For the lemma, I only show the unique existence of the experiments characterized by \cref{eq TC posterior}, \ref{eq TC reputation}, \ref{eq TC w}, and \ref{eq TC equality} and characterized by \cref{eq LB TC posterior}, \ref{eq LB TC reputation}, \ref{eq LB TC w}, and \ref{eq LB TC equality}. The reason why these equations characterize $\pi^\dagger$ and $\rho^\dagger$ is standard and mentioned in the main text.    

For $\pi^\dagger$, substitute \cref{eq TC posterior}, \ref{eq TC reputation}, and \ref{eq TC w} into \cref{eq TC equality}. The goal is then proving the unique existence of $\mu^\dagger_{\Tilde{a}}\in(\mu_0,1]$ that satisfies the following equation:
\begin{equation}
    2\mu^\dagger_{\Tilde{a}}-1=\theta\left[\frac{\tau}{1-(1-\tau)\mu_0(1-\mu^\dagger_{\Tilde{a}})/(1-\mu_0)\mu^\dagger_{\Tilde{a}}}-\frac{\tau}{\tau+(1-\tau)/\mu^\dagger_{\Tilde{a}}}\right].
\end{equation}
Define $H(\mu,\theta)$ as the difference between the left-hand side and the right-hand side, and $H(\mu^\dagger_{\Tilde{a}},\theta)=0$. Note that (1) $H(\cdot,\theta)$ is continuous and monotone in $\mu$, (2) $H(\mu_0,\theta)<0$, and (3) $H(1,\theta)>0$. Hence, for any $\theta$, there exists a unique $\mu^\dagger_{\Tilde{a}}\in(\mu_0,1)$ such that $H(\mu^\dagger_{\Tilde{a}},\theta)=0$. In particular, $\mu^\dagger_{\Tilde{a}}>1/2$ because $H(1/2,\theta)<0$. 

For $\rho^\dagger$, substitute \cref{eq LB TC posterior}, \ref{eq LB TC reputation}, and \ref{eq LB TC w} into \cref{eq LB TC equality}. The goal is then proving the unique existence of $\mu^\dagger_{\Tilde{b}}\in(0,\mu_0)$ that satisfies the following equation:
\begin{equation}
2\mu^\dagger_{\Tilde{b}}-1=\theta\left[\frac{\tau}{\tau+(1-\tau)/(1-\mu^\dagger_{\Tilde{b}})}-\frac{\tau}{1-(1-\tau)\mu^\dagger_{\Tilde{b}}(1-\mu_0)/\mu_0(1-\mu^\dagger_{\Tilde{b}})}\right].
\end{equation}
Define $K(\mu,\theta)$ as the difference between the left-hand side and the right-hand side, and $K(\mu^\dagger_{\Tilde{b}},\theta)=0$. Note that (1) $K(\cdot,\theta)$ is continuous and monotone in $\mu$, and (2) $K(0,\theta)<0$. Hence, for fixed $\theta$, there exists $\mu^\dagger_{\Tilde{b}}\in(0,\mu_0)$ such that $K(\mu^\dagger_{\Tilde{b}},\theta)=0$ if and only if $$K(\mu_0,\theta)=2\mu_0-1-\theta\left[\frac{\tau}{\tau+(1-\tau)/(1-\mu_0)}-1\right]>0.$$ Note that the term in the square bracket is negative. There exists $\underline{\theta}^\dagger>0$ such that $K(\mu_0,\theta)>0$ if and only if $\theta>\underline{\theta}^\dagger$. Consequently, if $\theta>\underline{\theta}^\dagger$, \cref{eq LB TC posterior}, \ref{eq LB TC reputation}, \ref{eq LB TC w}, and \cref{eq LB TC equality} admit an informative $\rho^\dagger$. Otherwise, $\rho^\dagger$ can be uninformative. \qed

\subsection{Proof of Proposition \texorpdfstring{\ref{prop TC}}{\ref*{prop TC}}}
To compare $\pi^{*}$ and $\pi^\dagger$, suppose $\pi^*=\pi^\dagger$. \cref{eq TC comparing w} is true because $\mu^*_{\Tilde{a}}=\mu^\dagger_{\Tilde{a}}>1/2$ and $w^\dagger(b)=1\geq w^\dagger(a)$. The reputation from correctly taking an action is then the same between \cref{eq basic reputation} and \cref{eq TC reputation}, i.e., $\tau^*(a,A)=\tau^\dagger(a,A)$ and $\tau^*(b,B)=\tau^\dagger(b,B)$. \cref{eq TC comparing reputation loss} can be proven as follows. Given $\pi^*=\pi^\dagger$ and $\mu^*_{\Tilde{a}}=\mu^\dagger_{\Tilde{a}}>1/2$,
\begin{equation}
    \begin{aligned}
        &[(1-\mu^*_{\Tilde{a}})\tau^*(b,B)-\mu^*_{\Tilde{a}}\tau^*(a,A)]-[w^\dagger(b)\tau^\dagger(b,B)-w^\dagger(a)\tau^\dagger(a,A)]\\
        =&\mu^*_{\Tilde{a}}\left[\frac{1-\mu^*_{\Tilde{a}}}{\mu^*_{\Tilde{a}}}\tau^*(b,B)-\tau^*(a,A)\right]-w^\dagger(a)\left[\frac{w^\dagger(b)}{w^\dagger(a)}\tau^\dagger(b,B)-\tau^\dagger(a,A)\right]\\
        &\text{(the term in the second square bracket is positive because of \cref{eq TC equality} and $\mu^\dagger_{\tilde{a}}>1/2$)}\\
        <&\mu^*_{\Tilde{a}}\left[\frac{1-\mu^*_{\Tilde{a}}}{\mu^*_{\Tilde{a}}}\tau^*(b,B)-\tau^*(a,A)\right]-\mu^*_{\Tilde{a}}\left[\frac{w^\dagger(b)}{w^\dagger(a)}\tau^\dagger(b,B)-\tau^\dagger(a,A)\right]\quad(\text{Note }\mu^*_{\tilde{a}}=\mu^\dagger_{\Tilde{a}}<w^\dagger(a))\\
        <&\mu^*_{\Tilde{a}}\left[\frac{1-\mu^*_{\Tilde{a}}}{\mu^*_{\Tilde{a}}}\tau^*(b,B)-\tau^*(a,A)\right]-\mu^*_{\Tilde{a}}\left[\frac{1-\mu^*_{\Tilde{a}}}{\mu^*_{\Tilde{a}}}\tau^\dagger(b,B)-\tau^\dagger(a,A)\right]=0.\\
    \end{aligned}
\end{equation}
The inequality together with the binding obedience constraint satisfied by $\pi^\dagger$ (\cref{eq TC equality}) implies that if $\pi^*=\pi^\dagger$, the obedience constraint is slacked as \cref{eq TC-C slack constraint}. 

Stick to the definition of $F(\mu,\theta)$ in \cref{proof lem equilibrium}. \cref{eq TC-C slack constraint} shows $F(\mu^\dagger_{\tilde{a}},\theta)>0$. Because $F(\mu^*_{\Tilde{a}},\theta)=0$, and $F(\mu,\theta)$ is monotone in $\mu$, we conclude $\mu^*_{\Tilde{a}}<\mu^{\dagger}_{\Tilde{a}}$. This fact with $\mu^*_{\Tilde{b}}=\mu^{\dagger}_{\Tilde{b}}=0$ shows $\pi^*<\pi^\dagger$. The transparency of decision consequences decreases the public's welfare because it decreases the information provision for the obedient politician.

For the comparison between $\rho^*$ and $\rho^\dagger$, recall that $\rho^*$ is characterized by \cref{eq LB basic posterior}, \ref{eq LB basic reputation}, \ref{eq LB basic equality}, the threshold $\underline{\theta}$, the prior $\mu_0$, and $\tau$. If $\tau\geq\mu_0/(1-\mu_0)$ or $\theta\leq\underline{\theta}$, $\rho^*$ is uninformative according to \cref{lem equilibrium}. It is then automatically true that $\rho^\dagger\geq\rho^*$. Below, assume $\tau<\mu_0/(1-\mu_0)$ and $\theta>\underline{\theta}$, and thus $\rho^*$ is informative.

Suppose $\rho^*=\rho^\dagger$. Because $\mu^*_{\Tilde{b}}=\mu^\dagger_{\Tilde{b}}<\mu_0<1/2$ and $w^\dagger(a)=1\geq w^\dagger(b)$, we have
$$\frac{1-\mu^*_{\Tilde{b}}}{w^\dagger(b)}>\frac{\mu^*_{\Tilde{b}}}{w^\dagger(a)},\text{ and}$$
\begin{equation}
\begin{aligned}
        &[(1-\mu^{*}_{\Tilde{b}})\tau^{*}(b,B)-\mu^{*}_{\Tilde{b}}\tau^{*}(a,A)]-[w^\dagger(b)\tau^\dagger(b,B)-w^\dagger(a)\tau^\dagger(a,A)]\\
        =&\mu^{*}_{\Tilde{b}}\left[\frac{1-\mu^{*}_{\Tilde{b}}}{\mu^{*}_{\Tilde{b}}}\tau^{*}(b,B)-\tau^{*}(a,A)\right]-w^\dagger(a)\left[\frac{w^\dagger(b)}{w^\dagger(a)}\tau^\dagger(b,B)-\tau^\dagger(a,A)\right]\\
        &\text{(the term in the second square bracket is negative because of \cref{eq LB TC equality} and $\mu^\dagger_{\Tilde{b}}<1/2$)}\\
\end{aligned}
\end{equation}
\begin{equation}
\begin{aligned}
        >&\mu^{*}_{\Tilde{b}}\left[\frac{1-\mu^{*}_{\Tilde{b}}}{\mu^{*}_{\Tilde{b}}}\tau^{*}(b,B)-\tau^{*}(a,A)\right]-\mu^{*}_{\Tilde{b}}\left[\frac{w^\dagger(b)}{w^\dagger(a)}\tau^\dagger(b,B)-\tau^\dagger(a,A)\right]\quad(\text{Note }\mu^*_{\tilde{b}}=\mu^\dagger_{\Tilde{b}}<w^\dagger(a))\\
        >&\mu^{*}_{\Tilde{b}}\left[\frac{1-\mu^{*}_{\Tilde{b}}}{\mu^{*}_{\Tilde{b}}}\tau^{*}(b,B)-\tau^{*}(a,A)\right]-\mu^{*}_{\Tilde{b}}\left[\frac{1-\mu^{*}_{\Tilde{b}}}{\mu^{*}_{\Tilde{b}}}\tau^\dagger(b,B)-\tau^\dagger(a,A)\right]=0.
\end{aligned}
\end{equation}
Stick to the definition of $L(\mu,\theta)$ in \cref{proof lem equilibrium}. The above inequalities together with the binding obedience constraint satisfied by $\rho^\dagger$ (\cref{eq LB TC equality}) imply that if $\rho^*=\rho^\dagger$,
\begin{equation}
    2\mu^*_{\Tilde{b}}-1=\theta[w^\dagger(b)\tau^\dagger(b,B)-w^\dagger(a)\tau^\dagger(a,A)]<\theta[(1-\mu^{*}_{\Tilde{b}})\tau^{*}(b,B)-\mu^{*}_{\Tilde{b}}\tau^{*}(a,A)],
\end{equation}
i.e., $L(\mu^\dagger_{\Tilde{b}},\theta)<0$. Because $L(\mu^*_{\Tilde{b}},\theta)=0$, and $L(\mu,\theta)$ is monotone in $\mu$, we conclude $\mu^*_{\Tilde{b}}>\mu^\dagger_{\Tilde{b}}$. This fact with $\mu^*_{\Tilde{a}}=\mu^\dagger_{\Tilde{a}}=1$ shows $\rho^*<\rho^\dagger$. The transparency of decision consequences decreases the public's welfare because it decreases the information provision for the obedient politician.\qed

\subsection{Proof of Proposition \texorpdfstring{\ref{prop TC 2}}{\ref*{prop TC 2}}}
The proposition is proven following three steps. In the proof, the essentially uninformative experiments are taken as uninformative to facilitate comparative statics. \medskip

\noindent\textbf{STEP 1:} Both $\pi^{**}$ and $\pi^\ddagger$ are informative, and $\rho^{**}$ and $\rho^\ddagger$ are uninformative.

Because for any $\alpha\in(0,1)$
$$\frac{(1-\mu_0)\tau}{1-\mu_0\tau}-1<(1-\mu_0)\tau-\mu_0<\frac{(1-\mu_0)\tau}{\tau+(1-\tau)(1-\alpha)}-\frac{\mu_0\tau}{\tau+(1-\tau)\alpha},$$
we can infer from \cref{lem experiments nontransparent preferences} and \cref{as TC} that $\pi^{**}$ is informative, and $\rho^{**}$ is uninformative. Similar to the proof of \cref{lem experiments nontransparent preferences}, if $\pi^\ddagger$ is unique and informative, $\rho^\ddagger$ must be uninformative. Hence, it suffices to prove that $\pi^\ddagger$ is unique and informative, which is characterized by \cref{eq TC 2 posteriors}, \ref{eq TC 2 reputations}, \ref{eq TC 2 w}, and \ref{eq TC 2 equality}. 

Substitute \cref{eq TC 2 posteriors}, \ref{eq TC 2 reputations}, \ref{eq TC 2 w} into \cref{eq TC 2 equality}. The binding obedience constraint can be written as
\begin{equation}\label{eq proof TC 2 reduced euqation}
    2\mu^\ddagger_{\Tilde{a}}-1=\theta\left[\frac{\tau}{\tau+(1-\tau)\left(1-\alpha\frac{\mu_0(1-\mu^\ddagger_{\Tilde{a}})}{(1-\mu_0)\mu^\ddagger_{\Tilde{a}}}+(1-\alpha)\frac{\mu_0}{1-\mu_0}\right)}-\frac{\tau}{\tau+(1-\tau)\alpha\frac{1}{\mu^\ddagger_{\Tilde{a}}}}\right].
\end{equation}
Define $J(\mu,\theta)$ as the difference between the left-hand side and the right-hand side, and $J(\mu^\ddagger_{\Tilde{a}},\theta)=0$. Note that (1) $J(\cdot,\theta)$ is continuous and monotone in $\mu$, and (2) $J(1,\theta)>0$. Then, $\mu^\ddagger_{\Tilde{a}}\in(\mu_0,1)$ exists if and only if $J(\mu_0,\theta)<0$. Notice that
$$
    J(\mu_0,\theta)
    \leq 2\mu_0-1+\theta\left[1-\frac{\tau(1-\mu_0)}{1-\mu_0\tau}\right]<0
$$
due to \cref{as TC}. Hence, there exists a unique $\mu^\ddagger_{\Tilde{a}}\in(\mu_0,1)$ such that $J(\mu^\ddagger_{\Tilde{a}},\theta)=0$, implying that $\pi^\ddagger$ is unique and informative.\medskip

\noindent\textbf{STEP 2:} There exists $\underline{\alpha}\in(0,1)$ such that $\mu^\ddagger_{\Tilde{a}}<1/2$ and $\mu^\ddagger_{\Tilde{a}}/w^\ddagger(a)<(1-\mu^\ddagger_{\Tilde{a}})/w^\ddagger(b)$.

Note 
\begin{equation}
    J\left(\frac{1}{2},\theta\right)=\theta\tau\left[\frac{1}{\tau+(1-\tau)2\alpha}-\frac{1}{\tau+(1-\tau)(1-\mu_0(2\alpha-1)/(1-\mu_0))}\right].
\end{equation}
The value is positive if $\alpha<1/2$. Because $J(\mu^\ddagger_{\Tilde{a}},\theta)=0$, and $J(\cdot,\theta)$ is monotone in $\mu$, we conclude $\mu^\ddagger_{\Tilde{a}}<1/2$ if $\alpha<1/2$. On the other hand, $\lim_{\alpha\to 0+}w^\ddagger(a)=1$ and $\lim_{\alpha\to 0+}w^\ddagger(b)<1$ in \cref{eq TC 2 w}. Consequently, there exists $\underline{\alpha}\in(0,1/2)$ such that if $\alpha<\underline{\alpha}$, $\mu^\ddagger_{\Tilde{a}}/w^\ddagger(a)<(1-\mu^\ddagger_{\Tilde{a}})/w^\ddagger(b)$, and \cref{eq TC 2 comparing w} holds.\medskip

\noindent\textbf{STEP 3:} If $\mu^\ddagger_{\Tilde{a}}<1/2$ and $\mu^\ddagger_{\Tilde{a}}/w^\ddagger(a)<(1-\mu^\ddagger_{\Tilde{a}})/w^\ddagger(b)$, $\pi^{**}>\pi^\ddagger$.

Suppose $\pi^{**}=\pi^\ddagger$. Then, $\tau^{**}(a,A)=\tau^\ddagger(a,A)$, $\tau^{**}(b,B)=\tau^\ddagger(b,B)$, and $\mu^{**}_{\Tilde{a}}=\mu^\ddagger_{\Tilde{a}}$. \cref{eq TC 2 change in reputational payoffs} can be proven as follows:
\begin{equation}
    \begin{aligned}
    &w^\ddagger(b)\tau^\ddagger(b,B)-w^\ddagger(a)\tau^\ddagger(a,A)-[(1-\mu^{**}_{\Tilde{a}})\tau^{**}(b,B)-\mu^{**}_{\Tilde{a}}\tau^{**}(a,A)]
    \end{aligned}
\end{equation}
\begin{equation}
    \begin{aligned}
    =&w^\ddagger(a)\left[\frac{w^\ddagger(b)}{w^\ddagger(a)}\tau^\ddagger(b,B)-\tau^\ddagger(a,A)\right]-\mu^{**}_{\Tilde{a}}\left[\frac{1-\mu^{**}_{\Tilde{a}}}{\mu^{**}_{\Tilde{a}}}\tau^{**}(b,B)-\tau^{**}(a,A)\right]\\
    &\text{(the term in the first square bracket is negative because of \cref{eq TC 2 equality} and $\mu^\ddagger_{\Tilde{a}}<1/2$)}\\
    <&\mu^\ddagger_{\tilde{a}}\left[\frac{w^\ddagger(b)}{w^\ddagger(a)}\tau^\ddagger(b,B)-\tau^\ddagger(a,A)\right]-\mu^{**}_{\Tilde{a}}\left[\frac{1-\mu^{**}_{\Tilde{a}}}{\mu^{**}_{\Tilde{a}}}\tau^{**}(b,B)-\tau^{**}(a,A)\right]\quad(\text{Note }\mu^\ddagger_{\Tilde{a}}<w^\ddagger(a))\\
    <&\mu^\ddagger_{\Tilde{a}}\left[\frac{1-\mu^\ddagger_{\Tilde{a}}}{\mu^\ddagger_{\Tilde{a}}}\tau^\ddagger(b,B)-\tau^\ddagger(a,A)\right]-\mu^{**}_{\Tilde{a}}\left[\frac{1-\mu^{**}_{\Tilde{a}}}{\mu^{**}_{\Tilde{a}}}\tau^{**}(b,B)-\tau^{**}(a,A)\right]=0.
    \end{aligned}
\end{equation}
The inequality and the binding obedience constraint satisfied by $\pi^\ddagger$ (\cref{eq TC 2 equality}) imply that if $\pi^{**}=\pi^\ddagger$, the experiment does not satisfy the obedience constraint, and \cref{eq C-TC stringent} holds. Then, for the function $G(\mu,\theta)$ defined in \cref{proof lem experiments nontransparent preferences}, we have $G(\mu^\ddagger_{\Tilde{a}},\theta)<0$. Because $G(\mu^{**}_{\Tilde{a}},\theta)=0$, and $G(\cdot,\theta)$ is monotone in $\mu$, we conclude $\mu^{**}_{\Tilde{a}}>\mu^\ddagger_{\Tilde{a}}$. This fact with $\mu^{**}_{\Tilde{b}}=\mu^\ddagger_{\Tilde{b}}=0$ proves $\pi^{**}>\pi^\ddagger$. The transparency of decision consequences increases the public's welfare because it increases the information provision for the obedient politician.\qed

\end{document}